\documentclass[%
reprint,
superscriptaddress,
amsmath,amssymb,
aps,
]{revtex4-1}

\usepackage{mathtools}
\usepackage{dcolumn}
\usepackage{bm}

\usepackage[dvips]{graphicx}
\usepackage{amsmath,amssymb,amsthm,mathrsfs,amsfonts,dsfont,amscd,keyval}
\usepackage{mathtools,mathrsfs}
\usepackage{yfonts} 
\usepackage{ytableau}
\usepackage{subfigure, epsfig}
\usepackage{braket}
\usepackage{bm}
\usepackage{fancyhdr}
\usepackage{enumerate}
\usepackage{color}
\usepackage{hyperref}
\usepackage{tabularx}
\usepackage{times}
\hypersetup{colorlinks=true, linkcolor=blue, citecolor=blue, urlcolor=black}
\usepackage{multirow}
\usepackage{float}
\usepackage{pdfpages}
\usepackage{pgffor}
\renewcommand\vec{\mathbf}

\theoremstyle{definition}
\newtheorem{definition}{Definition}[section]

\newtheorem{proposition}[definition]{Proposition}

\theoremstyle{plain}

\theoremstyle{remark}

\newcommand{\comments}[1]{}

\makeatletter
\AtBeginDocument{\let\LS@rot\@undefined}
\makeatother

\begin{document}

\title{Fundamental causal bounds of quantum random access memories}
	
\author{Yunfei Wang}

\affiliation{Martin A. Fisher School of Physics, Brandeis University, Waltham, MA 02453, USA}

\author{Yuri Alexeev}

\affiliation{Computational Science Division, Argonne National Laboratory, Lemont, IL 60439, USA}

\affiliation{Chicago Quantum Exchange, Chicago, IL 60637, USA}

\affiliation{Department of Computer Science, The University of Chicago, Chicago, IL 60637, USA}

\author{Liang Jiang}

\affiliation{Chicago Quantum Exchange, Chicago, IL 60637, USA}

\affiliation{Pritzker School of Molecular Engineering, The University of Chicago, Chicago, IL 60637, USA}

\author{Frederic T. Chong}

\affiliation{Department of Computer Science, The University of Chicago, Chicago, IL 60637, USA}

\author{Junyu Liu}

\affiliation{Chicago Quantum Exchange, Chicago, IL 60637, USA}

\affiliation{Department of Computer Science, The University of Chicago, Chicago, IL 60637, USA}

\affiliation{Pritzker School of Molecular Engineering, The University of Chicago, Chicago, IL 60637, USA}

\affiliation{Kadanoff Center for Theoretical Physics, The University of Chicago, Chicago, IL 60637, USA}

\affiliation{qBraid Co., Chicago, IL 60615, USA}

\affiliation{SeQure, Chicago, IL 60615, USA}
	
\date{\today}

\begin{abstract}
    Quantum devices should operate in adherence to quantum physics principles. Quantum random access memory (QRAM), a fundamental component of many essential quantum algorithms for tasks such as linear algebra, data search, and machine learning, is often proposed to offer $\mathcal{O}(\log N)$ circuit depth for $\mathcal{O}(N)$ data size, given $N$ qubits. However, this claim appears to breach the principle of relativity when dealing with a large number of qubits in quantum materials interacting locally. In our study we critically explore the intrinsic bounds of rapid quantum memories based on causality, employing the relativistic quantum field theory and Lieb--Robinson bounds in quantum many-body systems. In this paper, we consider a hardware-efficient QRAM design in hybrid quantum acoustic systems. Assuming clock cycle times of approximately $10^{-3}$ seconds and a lattice spacing of about 1 micrometer, we show that QRAM can accommodate up to $\mathcal{O}(10^7)$ logical qubits in 1 dimension, $\mathcal{O}(10^{15})$ to $\mathcal{O}(10^{20})$ in various 2D architectures, and $\mathcal{O}(10^{24})$ in 3 dimensions. We contend that this causality bound broadly applies to other quantum hardware systems. Our findings highlight the impact of fundamental quantum physics constraints on the long-term performance of quantum computing applications in data science and suggest potential quantum memory designs for performance enhancement.
\end{abstract}

\maketitle


\section{Introduction}
Quantum algorithms executed on quantum computers have the potential to provide significant benefits over their classical equivalents \cite{shor1999polynomial,grover1997quantum,harrow2009quantum}. However, a large portion of these algorithms necessitate data inputs, which are predominantly classical in today's computing sector. Consequently, the creation of interfaces that bridge classical and quantum processors may be a crucial factor in achieving broad and applicable quantum advantages. To illustrate, for an algorithm that classically handles $N$ data in polynomial time, we would need to transfer the classical data at an exponentially fast rate in $N$ to ensure that the overall quantum speedup is exponential. The development of such a fast quantum memory is a critical component for the future of the quantum industry, akin to the quantum version of data centers that merge quantum networks and quantum memories \cite{Liu:2022ubu}.

Quantum random access memory (QRAM) is a promising option to establish this efficient interface. For classical data of size $N$, QRAM circuits can achieve the upload of classical data within $\mathcal{O}(\log N)$ circuit depth and $\mathcal{O}(N)$ qubits. We primarily focus on the Bucket Brigade QRAM design due to its favorable circuit depth scaling. This design allows QRAM to serve as a fast quantum memory, as the coherence time of qubits does not scale rapidly with the qubit number $N$ and does not heavily rely on quantum information propagation within the lattice. Hence, our main concern lies in the bounds of from speed propagations. Numerous quantum algorithms, such as the Harrow--Hassidim--Lloyd algorithm used for linear systems of equations \cite{harrow2009quantum}, may depend on QRAM to ensure significant accelerations when dealing with dense classical data. Indeed, QRAM is also a fundamental component in many quantum machine learning applications \cite{Liu:2023coc,Liu:2022rxb,Jiang:2022oie}. However, the implementation of large-scale, fault-tolerant QRAM presents a substantial challenge in quantum hardware \cite{di2019methods,connorthesis,hann2019hardware,di2020fault,hann2021resilience}. While encouraging endeavors have been made in the development of QRAM, such as spin photon-based \cite{chen2021scalable} and hybrid quantum acoustic systems \cite{hann2019hardware}, there still exists ambiguity surrounding the timeline and methodology for large-scale QRAM production. We highlight that one potential hindrance for QRAM at an asymptotic scale is intricately tied to core principles in quantum physics: relativity and causality.

Quantum hardware is realized by quantum many-body systems, whose low-energy descriptions are Lorentz-invariant, local, relativistic quantum field theories. There is no up-to-date evidence about superluminal information transfer and significant Lorentz violation \cite{Adams:2006sv,Mattingly:2005re}. Assuming that interactions are local, it seems that fundamentally one cannot ensure an $\mathcal{O} (\log N)$ circuit depth for $\mathcal{O}(N)$ qubits asymptotically. Here we make a simple estimation with the following two assumptions. Suppose we place qubits as particles split by the lattice spacing $a =10^{-6} \text{m}$ in one dimension. The total length of the qubit chain is thus given by $L = a N$, where $N$ is the number of qubits. We further assume that running the total QRAM circuit takes time $T=\Delta T \log N$, where $\Delta T = 10^{-3} \text{s}$ per clock cycle. From causality, for 1 dimension chain QRAM system, we simply have
\begin{align} \label{Naive}
\frac{L}{T} = \frac{{Na}}{{\Delta T\log N}} \le c~,
\end{align}
where $c$ is the speed of light. Taking $c = 3 \times 10^8 \text{ m/s}$, we get an upper bound on the number of qubits $N \le 8.9\times 10^{12}$. That is, in the above setup, we can maximally have around 1000 billion qubits to deal with 1000 billion data in total.
The number itself is comparable to the numbers of training parameters and training date size in a commercial large language model in machine learning. For instance, \texttt{GPT-3} from OpenAI \cite{brown2020language} uses 175 billion training parameters to train 45,000 billion bytes of text data. If a quantum algorithm is somehow serving as part of the service with QRAM, it is approaching the bounds where causality is allowed to have exponential speedup \cite{Liu:2023coc}. The causality bound we describe is, as far as we know, original for generic quantum hardware. 

The objective of this paper is to substantiate and provide a rigorous framework for the statement above. Causality constraints in quantum many-body physics can be derived from the Lieb--Robinson bound for a local Hamiltonian \cite{lieb2004finite,Roberts:2016wdl,cramer2008locality}. In the low-energy description, basic gates in QRAM circuits can be represented as local interactions within relativistic quantum field theory. In our study we utilize  both approaches as examples to depict maximum information transmission using commutators in hybrid quantum acoustic systems that actualize QRAM \cite{hann2019hardware}. Not only will our work establish robust foundations for causal restrictions in QRAM, but it will also offer practical advice on quantum hardware design to enhance performance against causality bounds (refer to FIG.~\ref{FundamentalLimit3} and  FIG.~\ref{FundamentalLimit4}). Furthermore, our exploration of quantum gates as Feynman diagrams in relativistic quantum field theories (see FIG.~\ref{FDQG}) will contribute to our understanding of quantum hardware under extreme conditions. 

We will make two extra comments before we start detailed discussions. First, in the derivation of the bound Eqn.~(\ref{Naive}), the fundamental time scale $\Delta T$ is considered to be the clock cycle time for implementing QRAM circuits. In the actual hardware designs, this clock cycle time might be related or bounded by the gate operation time or the decoherence time of quantum hardware, representing operational time costs of each clock cycle. One can note that a smaller $\Delta T$ might lead to stronger bounds in $N$, since a smaller time separation might lead to smaller space separation for the same velocity. However, this does not indicate that we should not improve the gate time. For instance, a larger choice of $\Delta T$ will lead to higher infidelities for QRAM \cite{hann2021resilience}. Thus, a comprehensive design of QRAM hardware will lead to hybrid considerations including both causality bounds and fidelity. Second, we will primarily use the Hamiltonian of transmon and phonon in Eqn.~(\ref{QRAMEffHamiltonian}), where the speed limit is from the speed of sound determined from the hardware. However, one could also use classical communication, or alternative Hamiltonian constructions (and their corresponding quantum field theories ), and the speed limit could be given by the speed of light. To introduce classical communication, corresponding classical fields will be considered to enable the QRAM coupling, and could be treated as classical field backgrounds in the quantum systems (Thus, classical communication is a way to improve the causality bound from sound speed to the light speed, consistent with \cite{xu2023systems}). Thus, our results will be generic for known quantum materials following causality, considering the speed of light as the upper bound. Both of the issues will be discussed in more details in the Supplemental Materials.

Our paper is structured as follows. In Section \ref{sec:hardware} we will discuss our approach to QRAM hardware models within the context of hybrid quantum acoustic systems. Section~\ref{sec:qft} will provide a comprehensive discussion of causality bounds utilizing quantum field theory and Lieb--Robinson bounds, along with their implications for quantum hardware design. In Section~\ref{sec:conc} we will conclude and offer future perspectives. Some detailed proofs and technical discussions are presented in the Supplemental Materials. 

\section{QRAM hardware}\label{sec:hardware}

Our study employs QRAM hardware, the technical details of which have been meticulously outlined in references \cite{connorthesis,hann2019hardware}. We direct interested readers to these sources for a comprehensive analysis. In this section we offer a succinct overview of the procedure used for implementing quantum gates to extract data from a QRAM system comprising $N$ qubits, and we compute the total time of operation. The Hamiltonian that controls this QRAM system is as follows:

\begin{equation}\label{QRAMEffHamiltonian}
    \begin{split}
        & H = \omega_q q^{\dagger} q - \frac{\alpha}{2} q^{\dagger} q^{\dagger} q q\\
        & + \sum_k \bigg( \omega_k m_k^{\dagger} m_k + g_k q^{\dagger} m_k + g_k^{*} q m_k^{\dagger} \bigg)\\
        & + \sum_j \bigg( \Omega_j q^{\dagger} e^{i \omega_j t} + \text{H.C.} \bigg)~.
    \end{split}
\end{equation}

Here $q$ represents the annihilation operator for the transmon qubit, while $m_k$ represents the annihilation operator for the $k^{th}$ phonon mode. The transmon is described as an anharmonic oscillator with a Kerr nonlinearity parameter $\alpha$, and it is coupled to the $k^{th}$ phonon mode with a coupling strength of $g_k$. The external drives applied to the transmon with frequencies $\omega_j$, which are represented as $H_d = \sum_j \Omega_j q^{\dagger} e^{i \omega_j t} + \text{H.C.}$. In this content, $\Omega_j$ denotes the amplitude of the drive for the $j^{th}$ frequency, and H.C. stands for the Hermitian conjugate.

\begin{figure}[!ht]
\centering
\includegraphics[width=0.45\textwidth]{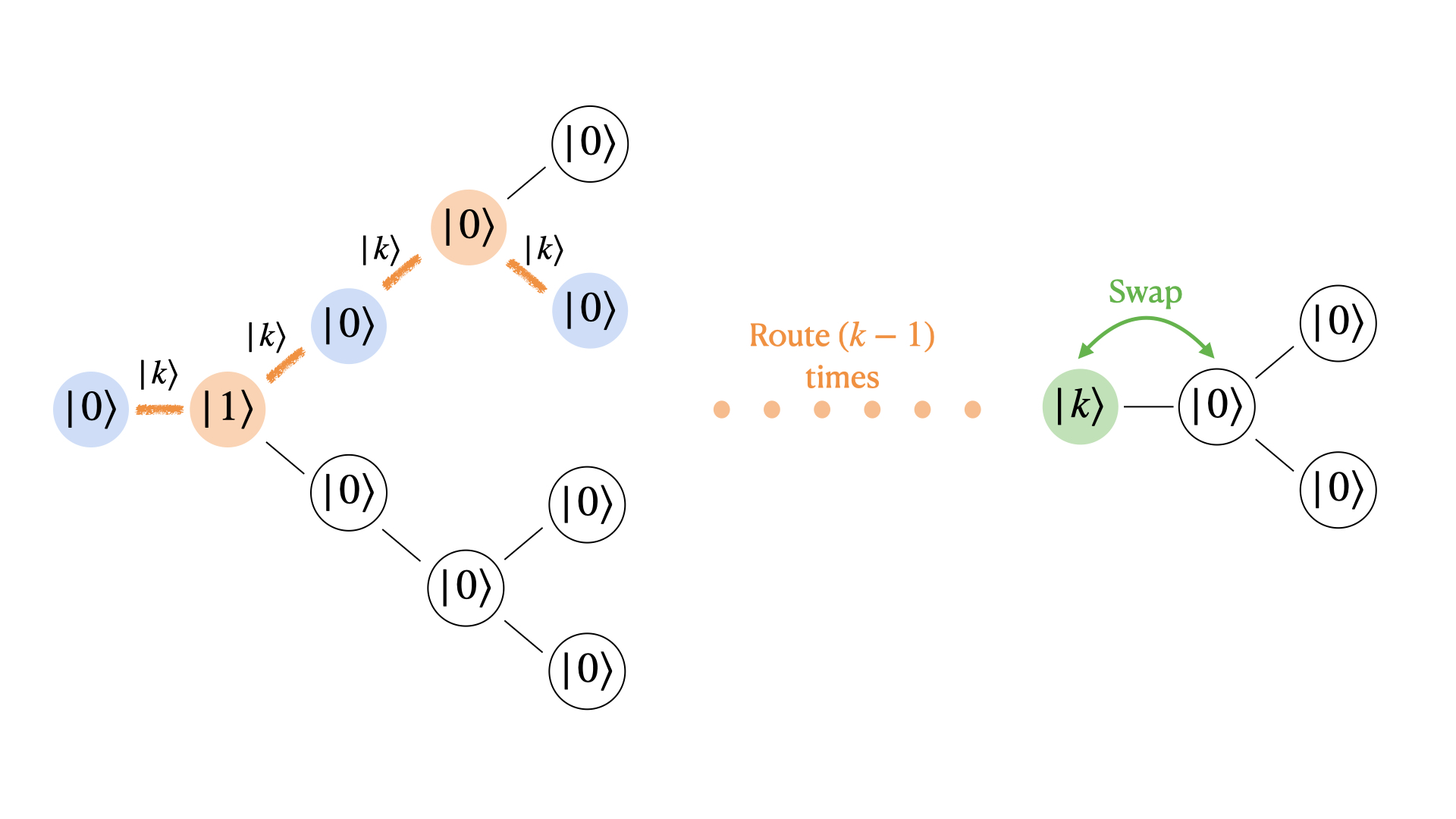}
\caption{During the $k$th step of initialization, the $k$th address qubit $\ket{k}$ follows the yellow branches in the figure, connecting the blue sites from left to right for each quantum router. $\ket{k}$ is then routed with a controlled-SWAP gate, which exchanges the qubit with the right channel, if the control qubit is in state $\ket{1}$, and exchanges with the left channel, if the control qubit is in state $\ket{0}$. The $\ket{k}$ qubit reaches the green site and a SWAP gate between $\ket{k}$, and the routing qubit will complete the initialization process.}
\label{kthStep}
\centering
\end{figure}

\begin{figure}[!ht] 
\centering
\includegraphics[width=0.45\textwidth]{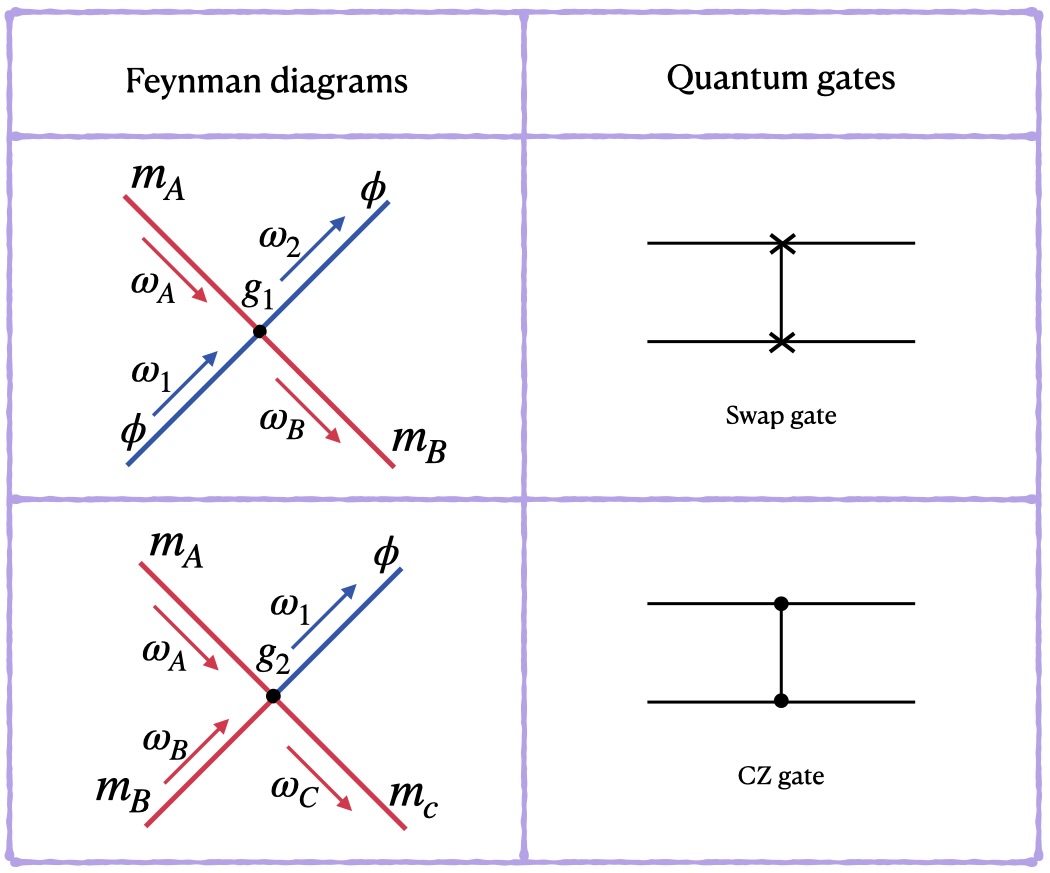}
\caption{Correspondence of Feynman diagrams of Eqn.~(\ref{EffectiveLagrangianDensity}) as quantum gates for realization of QRAM.}
\label{FDQG}
\centering
\end{figure}

The operation of the QRAM system is underpinned by two pivotal stages: initialization and routing. The initialization phase engages a SWAP gate, whereas the routing phase necessitates the integration of a controlled-SWAP gate and a SWAP gate. Throughout the initialization phase, the address qubit is systematically integrated into the system for individual initialization. For a QRAM comprising $N$ qubits, $\log{N}$ steps are required to accomplish comprehensive initialization. Each step, denoted as $k$ in FIG.~\ref{kthStep}, encompasses routing the $k$th address qubit for ($k-1$) times, followed by a SWAP operation to facilitate initialization.

Upon comprehensive initialization of the QRAM system, information extraction from the database becomes feasible. The precise extraction protocol is contingent on the type of information being accessed, as elaborated in \cite{hann2019hardware}. At the leading order, the overall operation times for extracting both classical and quantum information are essentially equivalent, with only insignificant discrepancies in numerical factors, where we use the time calculated in \cite{hann2019hardware}:

\begin{equation} \label{OperationTimeScale}
    T_{c \text{ or } q} \sim \tau_0 \times \log^2{N}~.
\end{equation}

Here we define $\frac{\pi}{g_1} + \frac{\pi}{g_2} \equiv \tau_0$. Unlike the linear dependence on $\log{N}$ in the total operation time of the simple example (Eqn.~(\ref{Naive})), the specific example we discuss in this paper exhibits a quadratic dependence on the circuit depth $\log{N}$.

\section{Lieb--Robinson bound, relativistic quantum field theory, and causality bounds}\label{sec:qft}

\subsection{Lieb--Robinson bound}

Considering the locality constraint inherent in QRAM,  employing a local Hamiltonian is essential. However, the one referenced in \cite{hann2019hardware} may not be ideally suited in scenarios around the speed of light (or sound) because of the unfettered coupling between the transmon qubit and phonon modes. As a result, we commence by formulating a local Hamiltonian that mirrors the system's properties:

\begin{equation} \label{FullDiscreteHamiltonian3D}
\begin{split}
    & H = H_{\mathcal{U}} + H_{\phi} + H_I\\
    = & \sum_{\vec{r}} \bigg( \frac{\mathcal{P}_{\vec{r}}^2}{2m} + \sum_{j}^{\nu} \sum_{\alpha,\beta} \frac{\lambda_j}{2} (\mathcal{U}_{\alpha}(\vec{r}) - \mathcal{U}_{\alpha}(\vec{r} + j \vec{e_{\beta}}))^2 \bigg)\\
    & + \int d^d \mathbf{x} \frac{1}{2} \bigg( \dot{\phi}(t,\mathbf{x})^2 + |\nabla \phi(t,\mathbf{x})|^2 \bigg)\\
    & + \int d^d \mathbf{x} \bigg[ \sum_{\vec{r}} \sum_{\alpha} \bigg( \mathcal{C} \phi(t,\mathbf{x}) \mathcal{U}_{\alpha}(\vec{r}) \bigg) + \frac{h}{4!} \phi^4(t,\mathbf{x}) \bigg]~.\\
\end{split}
\end{equation}

We consider the general case for $d = 1, 2, 3$ dimensions, where the indices $\alpha$ and $\beta$ refer to the directions in these dimensions. The lattice sites in the solid are denoted by the vector $\vec{r}$, and distortions in the $\alpha$ direction at site $\vec{r}$ are represented as $\mathcal{U}_{\alpha}(\vec{r})$. Additionally, $\vec{e}_{\beta}$ denotes the unit vector in the $\beta$ direction, and $\lambda_j$ represents the coupling constant between sites $\vec{r}$ and $\vec{r} + j \vec{e_{\beta}}$. The construction of $H_{\mathcal{U}}$ is detailed in the Supplemental Materials. $\phi$ is a massless scalar field, and $H_{\phi}$ corresponds to the standard Hamiltonian in quantum field theory textbooks \cite{Coleman:2018mew}. The interaction Hamiltonian is analogous to the interactions in Eqn.~(\ref{QRAMEffHamiltonian}), as explained also in the Supplemental Materials. 

For mathematical simplicity, throughout most of the paper we assume the isotropy of the solid. Otherwise, we would just have different frequency for modes traveling in different directions. As demonstrated in Eqn.~(\ref{FullDiscreteHamiltonian3D}), the coupling constants $\lambda_j$ remain the same for different values of $\alpha$. The construction of this Hamiltonian is fundamentally anchored in reintroducing locality into the system. We represent the phonon modes utilizing the position distortion operator, denoted as $\mathcal{U}_{\alpha}$. To confine couplings between distantly located sites, we impose a ceiling limit symbolized by $\nu$. This approach intuitively aligns with the concept that atoms distanced further apart are likely to be weakly coupled. Unlike the approach of modeling the transmon qubit as a single harmonic oscillator with a drive, as depicted in Eqn.~(\ref{QRAMEffHamiltonian}), we now portray it as a massless scalar field. Through transforming the drive frequency into the frequency of the $\phi$ operator, we further endorse locality, guided by the principles of special relativity and the architectural groundwork of quantum field theory. It is also important to note that we are working with a two-level system, hence qubits.

Thus, we present the Lieb--Robinson bound for the Hamiltonian given by Eqn.~(\ref{FullDiscreteHamiltonian3D}), which provides us with a bound on the velocity of information propagation. Define Weyl operator: $$W(f) \equiv \exp \bigg\{i \sum_{n \in \Lambda_L} \operatorname{Re}[f(n)]q_n + \operatorname{Im}[f(n)]p_n \bigg\}.$$ Here $f(n)$ is a complex function and $n$ denote sites in the whole system represented as $\Lambda_L$ here.

\begin{proposition}
    For Hamiltonian Eqn.~(\ref{FullDiscreteHamiltonian3D}), we have the following Lieb--Robinson bound for the operator commutator norm:

    \begin{equation}
        \begin{split}
            ||[\tau_t^{\Lambda} (W & (f)),  W(g)]|| \\
            \leq C \sum_{x,y \in \Lambda_L} & |f(x)| |g(y)|\\
            & e^{-\mu m \big[ d(X,Y) - c_{\omega,\lambda} \operatorname{max} \big( \frac{2}{\mu},e^{(\mu/2)+1} \big) |t| \big]}~,
        \end{split}
    \end{equation}

    where 

    \begin{equation}
        c_{\omega,\lambda} = \bigg( d \sum_{j}^{\nu} \frac{\lambda_{j}}{m} \bigg)^{1/2}~.
    \end{equation}
\end{proposition}

A meticulous and rigorous elaboration of the aforementioned causality statement is provided in the Supplemental Materials. Upon performing optimization with respect to $\mu$, we derive the subsequent Lieb--Robinson velocity:

\begin{equation} \label{dDSpeedLimit}
    v_h \leq 4 \bigg(d \sum_{j=1}^{\nu} \frac{\lambda_j}{m} \bigg)^{1/2}~.
\end{equation}

This bound acts as a maximum speed for transmitting information in a system described by Eqn.~(\ref{FullDiscreteHamiltonian3D}). Interestingly, the couplings to the transmon qubit do not affect this bound because they commute with the chosen Weyl operator. Additionally, Eqn.~(\ref{dDSpeedLimit}) is directly influenced by the parameter $\nu$, indicating that the presence of couplings between distant sites increases the speed limit. This is intuitive as stronger interactions between distant sites pose a challenge to locality. We note that the stronger the coupling between each site of the solid, the faster information can travel. Also noteworthy is the relationship between the speed limits in different dimensions (for $d = 1,2,3$):

\begin{equation} \label{2D3DSpeedLimitTo1D}
    \begin{split}
        v_h^{(d)} = \sqrt{d} v_h^{(1d)} = 4 \sqrt{d} \bigg(\sum_{j=1}^{\nu} \frac{\lambda_j}{m} \bigg)^{1/2}~.
    \end{split}
\end{equation}

Hence, a direct relationship exists between the material's dimensionality and the speed limit. Materials with higher dimensionality possess increased speed limits due to enhanced coupling opportunities and degrees of freedom. Furthermore, more pronounced dependencies are observable when considering the ratio between the number of qubits and the clock cycle time, a topic we will delve into in the subsequent sections.

\subsection{Causality bounds in relativistic quantum field theory} \label{CausalLim}

Alternatively, we propose a fresh perspective to comprehend the system by examining it in a low-temperature context. In this regime, only the long-wavelength modes with small $k$ are consequential. We implement a coarse-graining of the system by setting $k \rightarrow 0$ and subsequently perform a Legendre transformation to derive the subsequent Lagrangian. A comprehensive discussion on the coarse-graining procedure as well as the reason for the selection of this specific Lagrangian density can be found in the Supplemental Materials.

\begin{equation} \label{FullLagrangianDensitySimplified}
    \begin{split}
        \mathscr{L} = & \frac{\rho}{2} |\mathcal{\Dot{U}}|^2 - \sum_{\alpha,\beta}^{d} \frac{\lambda^{(d)}}{2}\bigg( \frac{\partial \mathcal{U}_{\alpha}}{\partial x^{\beta}} \bigg)^2 + \frac{1}{2} \partial_{\mu} \phi \partial^{\mu} \phi\\
        & - \sum_{\alpha} \mathcal{C} \phi \mathcal{U}_{\alpha} - \frac{h}{4!} \phi^4~.
    \end{split}
\end{equation}

As a result of the coarse-graining procedure, the site distortions $\mathcal{U}_{\alpha}(\vec{r})$ metamorphose into a continuous function $\mathcal{U}_{\alpha}(t, \vec{x})$. The coupling constant $\lambda^{(d)}$ in $d$ dimensions bears a direct correlation to the original couplings $\lambda_j$ in $d$ dimensions. By performing an expansion up to the first order in $k$, a relationship is established between the two coupling constants, as follows:

\begin{equation}\label{dDEquivalence}
    \lambda^{(d)} = d \sum_{j = 1}^{\nu} \lambda_j a j^2~.
\end{equation}

Our attention is devoted primarily to the causal bound of the system, a state attained via the development of the Lagrangian. By hypothesizing that the Lagrangian can be expressed as an integral of a density dependent on the derivatives of $\mathcal{U}_{\alpha}$, we can curtail interactions between distantly located sites. This is accomplished by analyzing the commutator $\big[ \pi_{\alpha}(x), \mathcal{U}_{\beta}(y) \big]$, an approach akin to the computation involved in the Lieb--Robinson bound. We initially underscore the fact that this quantity invariably amounts to zero when $x$ and $y$ are spacelike separated, a phenomenon attributed to Lorentz symmetry:

\begin{equation}
    \big[ \pi_{\alpha}(x), \mathcal{U}_{\beta}(y) \big] = 0 \hspace{3mm} \text{for $(x - y)^2 < 0$}~.
\end{equation}

This ensures that special relativity is respected and information cannot travel faster than the speed of light. To get more information, we focus on the analytical structure of $\big[ \pi_{\alpha}(x), \mathcal{U}_{\beta}(y) \big]$ when $x$ and $y$ are timelike separated, specifically for the case $d = 1$ as given in Eqn.~(\ref{LocalityConstraintSimplified}) up to leading order (for explicit calculation, see in the Supplemental Materials):

\begin{equation} \label{LocalityConstraintSimplified}
     \big[\mathcal{U}_{\beta}(y), \pi_{\alpha}(x) \big] < \frac{1}{\pi} \frac{e^{- (r - \sqrt{\lambda^{(1d)}/\rho} t) \epsilon}}{\big( r - \sqrt{\lambda^{(1d)}/\rho} t \big)}~.
\end{equation}

In 1D, the commutator is exponentially suppressed if $(r - \sqrt{\lambda^{(1d)}/\rho} t) > 0$. This sets the velocity limit for the system in 1D:

\begin{equation}
    v^{(1d)} = \sqrt{\lambda^{(1d)}/\rho}~.
\end{equation}

The velocity limits in different dimensions ($d = 1, 2, 3$) follow a  relation similar to that in the Lieb--Robinson bound:

\begin{equation} \label{dDSpeedLimit2}
    v^{(d)} = \bigg(\frac{\lambda^{(d)}}{\rho}\bigg)^{1/2} = \sqrt{d} v^{(1d)} = \sqrt{d} \bigg(\frac{\lambda^{(1d)}}{\rho}\bigg)^{1/2}~,
\end{equation}

It is not surprising that the bound is independent of $\phi$ and interaction terms, as the interaction parts commute with $\mathcal{U}_{\alpha}$. By relating the coupling constants in the discrete and continuous cases through coarse graining, we observe that the two speed limits are effectively the same. The numerical factor $j$ differs because of the imprecision of coarse graining when considering farther site couplings and ignoring higher derivative terms in Eqn.~(\ref{FullLagrangianDensitySimplified}). However, if we consider only couplings between nearest neighbors, the two speed limits are equal. This can be understood intuitively, as $\partial \mathcal{U}_{\alpha} / \partial x^{\beta}$ is the continuous counterpart of the coupling between nearest neighbors. Couplings between more distant sites can  be accurately described only by including higher derivatives of $\mathcal{U}_{\alpha}$ in the Lagrangian, which challenges our assumptions of locality.

\subsection{Scattering process as quantum gates}

The Feynman diagrams corresponding to Eqn.~(\ref{FullLagrangianDensitySimplified}) can reproduce the wave mixing process described in \cite{hann2019hardware}. By utilizing the Kerr nonlinearity in Eqn.~(\ref{QRAMEffHamiltonian}), the authors achieved wave mixing between external drives and phonon modes, resulting in the realization of quantum gates for QRAM.

We observe that the interaction term $\phi \mathcal{U}_{\alpha}$ allows us to convert any incoming $\phi$ field into a phonon mode in the Feynman diagrams generated by Eqn.~(\ref{FullLagrangianDensitySimplified}). By attaching $\mathcal{U}_{\alpha}$ to the tree-level diagram via $\phi^4$ interactions, we can mix the transmon waves and phonon modes, reproducing the wave mixing process described in \cite{hann2019hardware}.

Since $\phi$ is a scalar field, its propagator at the tree-level process, for dimensions $d = 1,2,3$, is a classical number. We can incorporate the propagators into the vertices and introduce effective coupling constants $g_1$ and $g_2$. Consequently, we construct the following effective Lagrangian density. We put $l_0$ in just to keep $g_1$ and $g_2$ dimension fixed for different $d$. (Details of this wave-mixing processes are presented in the Supplemental Materials.) 

\begin{equation} \label{EffectiveLagrangianDensity}
    \begin{split}
        \mathscr{L} = & \frac{\rho}{2} |\mathcal{\Dot{U}}|^2 - \sum_{\alpha,\beta}^{d} \frac{\lambda^{(d)}}{2}\bigg( \frac{\partial \mathcal{U}_{\alpha}}{\partial x^{\beta}} \bigg)^2\\
        & + \frac{1}{2} \partial_{\mu} \phi \partial^{\mu} \phi - \frac{g_1}{4 l_0^{2-d}} \sum_{\alpha} \phi^2 \mathcal{U}_{\alpha}^2\\
        & - \frac{g_1}{2 l_0^{2-d}} \sum_{\alpha \neq \beta} \phi^2 \mathcal{U}_{\alpha} \mathcal{U}_{\beta}  - \frac{g_2}{3! l_0^{2-d}} \sum_{\alpha} \phi \mathcal{U}_{\alpha}^3\\
        &  - \frac{g_2}{l_0^{2-d}} \sum_{\alpha \neq \beta \neq \gamma} \phi \mathcal{U}_{\alpha} \mathcal{U}_{\beta} \mathcal{U}_{\gamma} - \frac{g_2}{2 l_0^{2-d}} \sum_{\alpha \neq \beta} \phi \mathcal{U}_{\alpha}^2 \mathcal{U}_{\beta}~.
    \end{split}
\end{equation}

Hence, there exists a correspondence between the Feynman diagrams generated by Eqn.~(\ref{EffectiveLagrangianDensity}) and the quantum gates employed in QRAM.

Through direct calculations, as presented in the Supplemental Materials, we determine the time scale for the three gates that will be utilized. The time scale for the SWAP gates, beam-splitter, and CZ gates are determined respectively as follows: $t_{\text{sw}} = \frac{\pi}{2 g_1}$, $t_{\text{bs}} = \frac{\pi}{4 g_1}$, and $t_{\text{cz}} = \frac{\pi}{g_2}$. Importantly, these time scales are in the identical form as the times calculated in \cite{hann2019hardware}.

\begin{figure}[!ht] 
\centering
\includegraphics[width=0.45\textwidth]{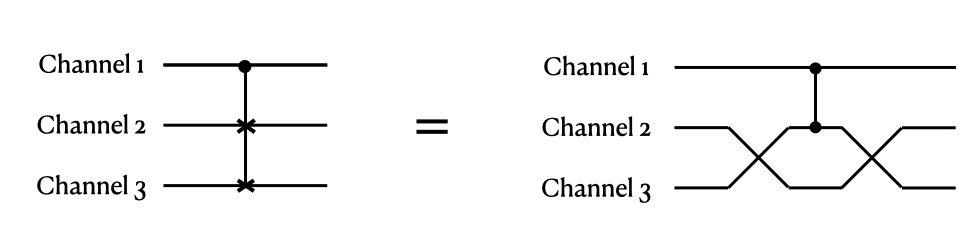}
\caption{Realization of the controlled-SWAP gate from basic gates: beam-splitter and CZ.}
\label{CSWAP}
\centering
\end{figure}

To construct QRAM, an essential building block is the controlled-SWAP gate, as depicted in FIG.~\ref{kthStep}. In our specific example, the controlled-SWAP gate is realized using a combination of basic gates: beam-splitter + CZ + beam-splitter, illustrated in FIG.~\ref{CSWAP}. 

The detailed demonstration of this construction can be found in the Supplemental Materials and the reference \cite{hann2019hardware}. It's important to note that the decomposition of controlled-SWAP gates shown in FIG.~\ref{CSWAP} is a specific case, assuming channel 1 is a qubit and channels 2 and 3 can be bosonic modes. Under the assumption that all modes are two-level systems, there might be even simpler decompositions of the controlled-SWAP gates, such as using 3 Toffoli gates.

Reference \cite{gao2018entangling} provides experimental demonstrations for the controlled-SWAP gates in the specific design shown in FIG.~\ref{CSWAP}. However, that paper did not demonstrate the coherence of the controlled-SWAP due to significant dephasing in channel 1. On the other hand, reference \cite{chapman2022high} uses a slightly different method and improved device coherence properties to demonstrate deterministic controlled-SWAP operation.

For classical communication systems, electromagnetic fields can be treated classically with interactions based on specific designs. The key is to identify the medium through which the information is carried. In the case of classical light speed communication, Maxwell's equations can be manipulated to derive the wave equation for electromagnetic waves, which reveals the speed limit of $c = 3 \times 10^8 \text{ m/s}$ for these waves. This result is well-documented in classical electromagnetic textbooks \cite{feynman2011feynman,jackson2021classical,griffiths2017introduction}. The upper bound of the memory system can be estimated by analyzing the specific design of the communication system, similar to previous sections.

\subsection{Limit of QRAM by locality}

The speed limits imposed by Eqn.~(\ref{dDSpeedLimit}) or Eqn.~(\ref{dDSpeedLimit2}) place constraints on the size of the QRAM. In a system governed by Hamiltonian Eqn.~(\ref{FullLagrangianDensitySimplified}), all information propagation is subject to these bounds, which are equivalent to the Lieb--Robinson bound of Hamiltonian Eqn.~(\ref{FullDiscreteHamiltonian3D}). The information stored in the database can travel at most a distance of $N \times a$ (where $N$ is the total number of qubits and $a$ is the separation distance) by applying a series of quantum gates generated by Eqn.~(\ref{FullLagrangianDensitySimplified}). Therefore, the total time required for the information to travel is given by the operation time calculated in Eqn.~(\ref{OperationTimeScale}). It is essential to ensure that the speed at which information travels is slower than the bound described by Eqn.~(\ref{dDSpeedLimit}) or Eqn.~(\ref{dDSpeedLimit2}).

\begin{equation}
    \begin{split}
        \frac{N}{\log^2{N}} \leq \frac{4 \sqrt{d} \big(\frac{\pi}{g_{1}} + \frac{\pi}{g_{2}} \big)}{a} \bigg(\sum_{j=1}^{\nu} \frac{\lambda_j}{m} \bigg)^{1/2} \hspace{2mm} \text{or} \hspace{2mm} \bigg(\frac{\lambda^{(1d)}}{\rho}\bigg)^{1/2}~.
    \end{split}
\end{equation}

\begin{figure}[!ht]
\centering
\includegraphics[width=0.45\textwidth]{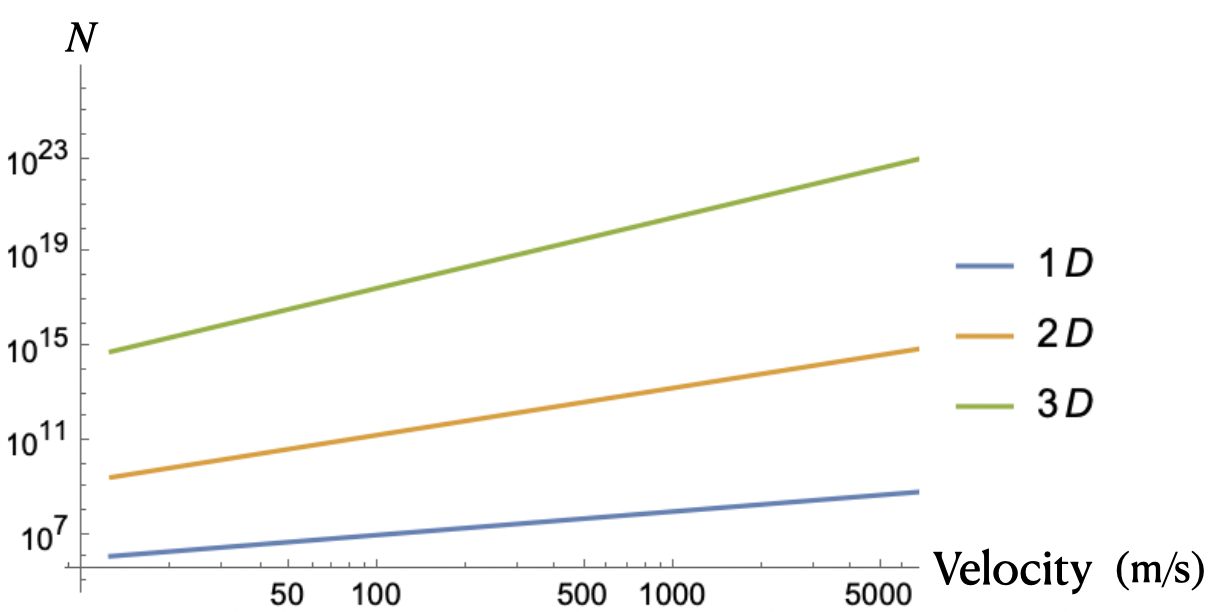}
\caption{Bounds of QRAM size $N$ for dimensions 1, 2, and 3. Bounds of QRAM size $N$ for dimensions 1, 2, and 3. Here we assume the lattice spacing of $10^{-6}~$m and the clock cycle time of $10^{-3}~$s. The horizontal axis is equal to the velocity limit determined by $\sum_{j=1}^\nu \sqrt{d} \left(\frac{\lambda_j}{m}\right)^{1/2}$ or $ \sqrt{\frac{\lambda^{(d)}}{\rho}}$. This velocity is taken to be at most on the order of typical sound speed in solids: about $6000~$m/s. }
\label{FundamentalLimit3}
\centering
\end{figure}

We illustrate the relationship between the material constants used in constructing QRAM and the size bound in FIG.~\ref{FundamentalLimit3}.

We assume that the clock cycle time used in FIG.~\ref{FundamentalLimit3} is $10^{-3}$ second. The identification of this is not important, since what we care about is the total operation time. In terms of dissecting the total operation time into either individual gates or clock cycles will not make a big difference. The rough $\log{N}$ scaling for time will work for both identifications. We also consider the speed limit to be approximately the sound speed in the resonator, which is around $6 \times 10^3 \text{ m/s}$. However, it is possible to exceed this bound with alternative QRAM designs, in which case the ultimate speed limit would be the speed of light, $c = 3 \times 10^8 \text{ m/s}$. 

The upper bounds for QRAM systems in higher dimensions are larger mainly because of the increased number of qubits they can accommodate (speed limits are also higher, as discussed before). In 2D systems, the total number of qubits is proportional to $N^2$ and $N^3$ for 3D. This difference in scaling allows for a significantly higher bounds in higher-dimensional hardware designs, as illustrated in FIG.~\ref{FundamentalLimit3}. The quadratic dependence of the total operation time on the circuit depth in this model does not significantly impact the overall results (at most up to one order of magnitude). It is specific to the model considered in this paper. For other models, one can calculate the corresponding dependence and extrapolate the results presented here.

Regarding differences between dimensions, we specifically mention an architecture design described in \cite{xu2023systems}, where quantum teleportation is used for qubit routing instead of SWAP actions. While most quantum gates in this design are confined by the sound speed of the 2D solid, the routing steps employing quantum teleportation are constrained by the speed of light. As the system size increases, a larger proportion of qubit distances are covered by quantum teleportation, resulting in only a few steps being confined by sound speed. Based on this observation, we estimate the maximum bounds for the QRAM size $N$ in this 2D design to be on the order of $\mathcal{O}(10^{20}) \sim \mathcal{O}(10^{22})$ qubits, whose performance is somewhere between 2D and 3D in FIG.~\ref{FundamentalLimit3}. For explicit calculations, please see the Supplemental Materials. 

In FIG.~\ref{FundamentalLimit4}, the horizontal axis represents the coupling constants, while the vertical axis represents the square of the speed limit. We consider $g_1$ and $g_2$ to have approximately the same value, both smaller than 1. This assumption is fundamental in perturbation theory, as used in \cite{hann2019hardware}, and in quantum field theory in our analysis.

\begin{figure}[!ht] 
\centering
\includegraphics[width=0.47\textwidth]{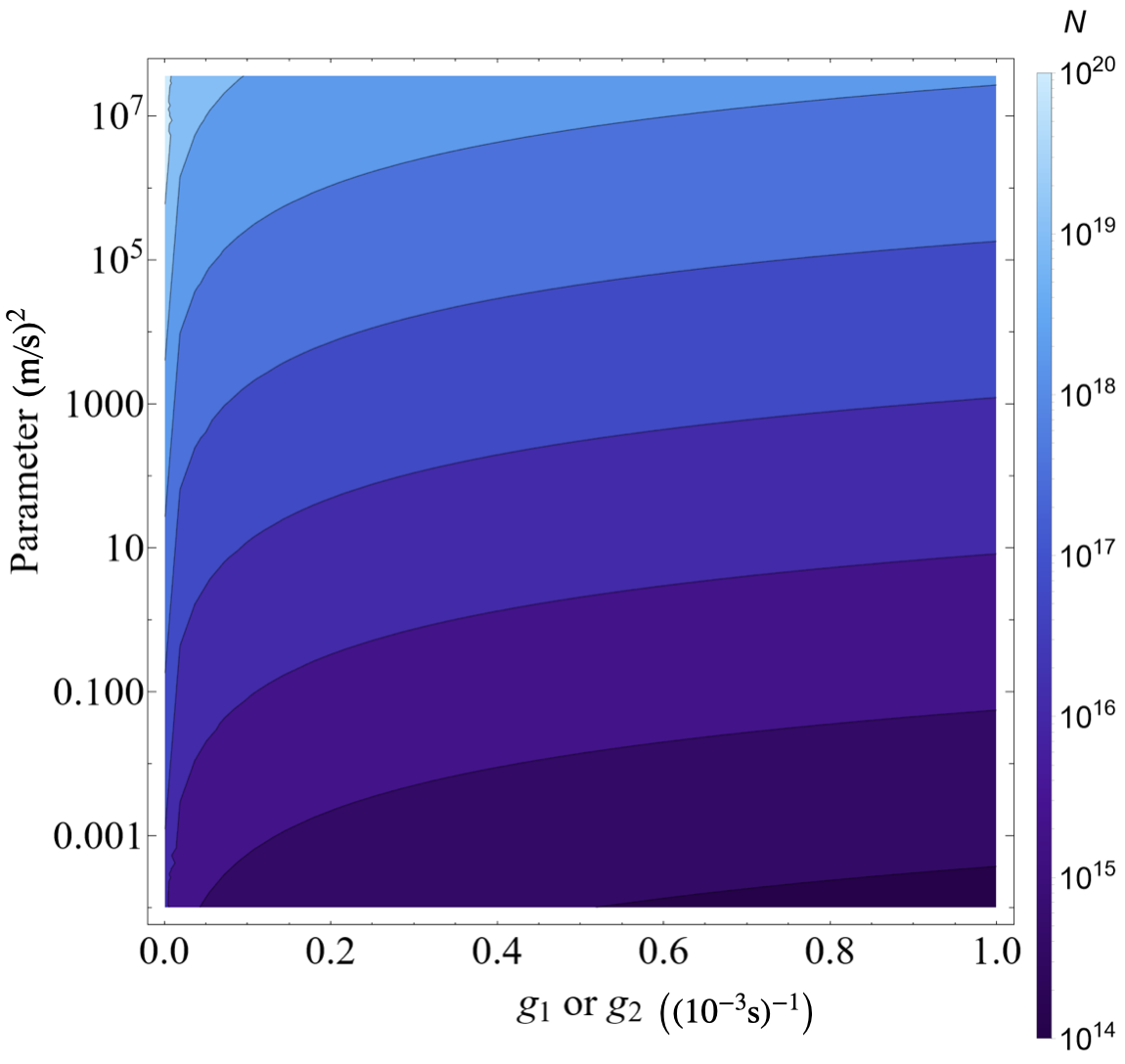}
\caption{Heat plot for bounds of 1 dimension QRAM size $N$. We also assume that the lattice spacing is $10^{-3}~$m. The speed limit is again on the order of the sound speed in solids. For parameters in the vertical axes we indicate $\sum_{j=1}^\nu \left(\frac{\lambda_j}{m}\right)$ or $\frac{\lambda^{(1d)}}{\rho}$.} 
\label{FundamentalLimit4}
\centering
\end{figure}

With more realistic clock cycle times, the restriction on the number of qubits, $N$, becomes even larger, extending to up to $10^{14}$ qubits. However, it is crucial not to excessively lean toward the left in FIG.~\ref{FundamentalLimit4}, as doing so would result in exceptionally prolonged operation times for QRAM. To elevate the upper bound on $N$, one would need to identify a solid where atoms are strongly coupled or, alternatively, decrease the coupling constants between the transmon qubit and the resonator to extend the clock cycle times. Nevertheless, the upper bounds depicted in FIG.~\ref{FundamentalLimit3} and FIG.~\ref{FundamentalLimit4} already suffice for the requirements of most quantum algorithms involving large quantities of qubits.

\section{Conclusion}\label{sec:conc}
In this paper we establish a universal bound on the maximum number of qubits permissible in quantum random access memory (QRAM), derived from causality and relativity principles. We substantiate the causality bounds through rigorous applications of relativistic quantum field theories and Lieb--Robinson bounds in quantum apparatus, using hybrid quantum acoustic systems as illustrative examples. While the bounds are elucidated by using specific examples, we propose that similar bounds would likely be present in general quantum devices constructed by local quantum many-body physics systems that implement QRAM. Our research not only offers essential insights into the constraints of quantum devices under extreme circumstances but also proposes strategies to optimize hardware designs, ensuring rapid quantum memory in the asymptotic context.

Our work initiates a new avenue of research that delves into the fundamental aspects of quantum memories and their computational constraints. Future research trajectories may encompass the development of strategies to optimize and circumvent causality bounds, comprehension of the ultimate capabilities of quantum memories for specific quantum algorithm applications, the conception of innovative alternative quantum hardware models with augmented connectivity and nonlocal interactions, and the exploration of implications on the causality bound from hybrid quantum-classical co-designs that take into account data contents. We present these potential research themes as fertile ground for future investigation.


\begin{acknowledgments}
We thank Scott Aaronson for helpful discussions about QRAM in NSF Workshop on Quantum Advantage and Next Steps in 2022. We thank Yongshan Ding, Jens Eisert and Shifan Xu for many helpful conversations. Y.A. acknowledges support from DOE Q-NEXT and DE-AC02-06CH11357. L.J. acknowledges support from the ARO(W911NF-23-1-0077), ARO MURI (W911NF-21-1-0325), AFOSR MURI (FA9550-19-1-0399, FA9550-21-1-0209), NSF (OMA-1936118, ERC-1941583, OMA-2137642), NTT Research, Packard Foundation (2020-71479), and the Marshall and Arlene Bennett Family Research Program. This material is based upon work supported by the U.S. Department of Energy, Office of Science, National Quantum Information Science Research Centers. This work is funded in part by EPiQC, an NSF Expedition in Computing, under award CCF-1730449; in part by STAQ under award NSF Phy-1818914; in part by NSF award 2110860; in part by the US Department of Energy Office of Advanced Scientific Computing Research, Accelerated Research for Quantum Computing Program; and in part by the NSF Quantum Leap Challenge Institute for Hybrid Quantum Architectures and Networks (NSF Award 2016136) and in part based upon work supported by the U.S. Department of Energy, Office of Science, National Quantum Information Science Research Centers. F.T.C. is Chief Scientist for Quantum Software at ColdQuanta and an advisor to Quantum Circuits, Inc. J. L. is supported in part by International Business Machines (IBM) Quantum through the Chicago Quantum Exchange, and the Pritzker School of Molecular Engineering at the University of Chicago through AFOSR MURI (FA9550-21-1-0209).
\end{acknowledgments}

\bibliographystyle{apsrev4-1}
\bibliography{QRAM.bib}

\newpage
\pagebreak
\clearpage
\foreach \x in {1,...,\the\pdflastximagepages}
{
	\clearpage
	

\section*{Supplementary material}

# Supplemental Materials: Fundamental causal bounds of quantum random access memories

## I. SPEED LIMIT FOR INFORMATION PROPAGATION

Quantum random access memory (QRAM) is a system for extracting information from a database. One of the most promising QRAM systems is described by the following Hamiltonian.

$$H = \omega_q q^\dagger q - \frac{\alpha}{2} q^\dagger q^\dagger q q + \sum_k \left( \omega_k m_k^\dagger m_k + g_k q^\dagger m_k + g_k^* q m_k^\dagger \right) + \sum_j \left( \Omega_j q^\dagger e^{i\omega_j t} + H.C. \right). \quad (1)$$

The first part describes the transmon qubit, which is modeled as a simple harmonic oscillator of frequency $\omega_q$ with a Kerr nonlinearity term. The second part refers to the phonon modes of the resonator, which are piezoelectrically coupled with the transmon qubit. The last line represents an external drive that can be applied to the transmon qubit. Note that this effective Hamiltonian is nonlocal because the transmon qubit is coupled to infinitely many phonon modes at once (there is no clear upper bound for $k$ in Eqn. (1)). One advantage of QRAM compared with its classical counterparts is that it can operate much faster because of superposition. See [1] for details on this matter.

Like any other system, however, a QRAM should be constrained by locality, which is ultimately implied in special relativity. Information cannot travel faster than the speed of light. In this section we analyze the speed limit for information propagation in a QRAM system equivalent to Eqn. (1). We need to construct a local Hamiltonian that reproduces quantum gates to realize QRAM, as Eqn. (1) did. We will make this explicit in a later section. The idea is to utilize the usual phonon Hamiltonian construction [2] but change the way we couple it with the transmon qubit compared with Eqn.(1), where we directly coupled the phonon modes with the transmon qubit through ladder operators. Here, we will couple them through field operators.

### A. Construction of phonon Hamiltonian

Consider a solid that has atoms at sites denoted by $\mathbf{r} = (l, m, n) = l\mathbf{a} + m\mathbf{b} + n\mathbf{c}$, where $\mathbf{a}$, $\mathbf{b}$, and $\mathbf{c}$ are vectors. The positions of the atoms are given by $\mathbf{q_r}$. The following construction is true for $d$ dimensions, where realistically we consider $d = 1, 2, 3$. When we input energy into the system or consider it at finite temperature, the atoms will start to move around, distorting the lattice and causing the position of atoms at a particular site to become $\mathbf{qr} = \mathbf{q_r} + \mathcal{U}_\mathbf{r}$. The most general Hamiltonian for this lattice is as follows:

$$H = \sum_{\mathbf{r}} \left( \frac{\mathbf{p}_{\mathbf{r}}^2}{2m} + \mathcal{V}(\mathbf{q}_{\mathbf{r}}) \right). \tag{2}$$

Note that we consider the system at a low temperature since our main focus is on the solid not melting and preserving its integrity. In this regime we can expand the potential around the lowest point, denoted as $\mathcal{V}_0$. Additionally, due to this expansion, all first derivative terms vanish.

$$\mathcal{V}(\mathbf{q}_{\mathbf{r}}) = \mathcal{V}_0 + \frac{1}{2} \sum_{\mathbf{r},\mathbf{r}'} \sum_{\alpha,\beta=1}^{d} \frac{\partial^2 \mathcal{V}}{\partial q_{\mathbf{r}}^\alpha \partial q_{\mathbf{r}'}^\beta} \mathcal{U}_{\mathbf{r}}^\alpha \mathcal{U}_{\mathbf{r}'}^\beta + \mathcal{O}(\mathcal{U}^3). \tag{3}$$

Furthermore, due to translation symmetry (the energy does not change if we move the entire solid to a different location), the term $\frac{\partial^2 \mathcal{V}}{\partial q_{\mathbf{r}}^\alpha \partial q_{\mathbf{r}'}^\beta}$ should not depend on the specific positions $\mathbf{r}$ and $\mathbf{r}'$ but rather on their relative position $\mathbf{r} - \mathbf{r}'$. As a result, we can model the Hamiltonian as follows:

$$H_{\mathcal{U}} = \sum_{\mathbf{r}} \frac{\mathcal{P}_{\mathbf{r}}^2}{2m} + \sum_{j}^{\nu} \sum_{\alpha} \frac{\lambda_j}{2} (\mathcal{U}_\alpha(\mathbf{r}) - \mathcal{U}_\alpha(\mathbf{r}'))^2. \tag{4}$$

We note that we have ignored the center of mass movement and zero-point energy, so we have $\mathcal{P}_\alpha(\mathbf{r}) = m\dot{\mathcal{U}}_\alpha(\mathbf{r})$. The upper bound $\nu$ is ultimately determined by the size of the system. However, we should also consider that the coupling between farther sites should intuitively be weaker. Therefore, we can terminate the sum on $j$ by setting $\nu$ as an upper bound before it reaches the boundary of the solid. We emphasize that Eqn. (4) represents a very general Hamiltonian. It includes numerous "cross" interactions, where $\mathbf{r} - \mathbf{r}' \neq j\mathbf{e}_\alpha$, and $\mathbf{e}_\alpha$ represents the unit vector in the $\alpha$ direction. For our purposes, we would like to simplify this Hamiltonian.

### B. Simplified phonon Hamiltonian

#### 1. Phonon Hamiltonian in 1D chain

We first consider a 1D chain with $n$ sites. The most general Hamiltonian in 1D that still satisfies the assumptions we have made above is as follows (in 1D, there are naturally no "cross" couplings):

$$H = \sum_{n} \frac{\mathcal{P}_n^2}{2m} + \sum_{j}^{\nu} \sum_{n} \frac{\lambda_j}{2} \left( \mathcal{U}_{n+j} - \mathcal{U}_n \right)^2. \tag{5}$$

The equation of motion is as follows:

$$m\ddot{\mathcal{U}}_n = -\sum_{j=1}^{\nu} \lambda_j (2\mathcal{U}_n - \mathcal{U}_{n+j} - \mathcal{U}_{n-j}). \tag{6}$$

The solutions to Eqn. (6) are plane wave functions. We gather them together to form the most general solution:

$$\mathcal{U}_n(t) = \sum_{l \neq 0} \left[ \alpha_l e^{-i(\omega_l t - k_l na)} + \alpha_l^\dagger e^{i(\omega_l t - k_l na)} \right]. \tag{7}$$

Plugging Eqn. (7) into the equation of motion, we obtain an important frequency spectrum of the system:

$$\omega_l(k) = \left( \sum_{j=1}^{\nu} \frac{\lambda_j}{m} (2 - 2\cos k_l ja) \right)^{1/2}. \tag{8}$$

Additionally, we have the canonical momentum:

$$\mathcal{P}_n(t) = m\dot{\mathcal{U}}_n = \sum_{l \neq 0}\left[-im\omega_l \alpha_l e^{-i(\omega_l t - k_l na)} + im\omega_l \alpha_l^\dagger e^{i(\omega_l t - k_l na)}\right]. \tag{9}$$

The commutation relation is given by $[\mathcal{U}_n, \mathcal{P}_m] = i\delta_{n,m}$. From this, we can deduce the expressions of the two "ladder" operators and their commutation relation:

$$\begin{aligned}\alpha_l &= \frac{1}{2m\omega_l N}\sum_n e^{ik_l na}\left(m\omega_l \mathcal{U}_n + i\mathcal{P}_n\right), \\ \alpha_l^\dagger &= \frac{1}{2m\omega_l N}\sum_n e^{ik_l na}\left(m\omega_l \mathcal{U}_n - i\mathcal{P}_n\right).\end{aligned} \tag{10}$$

Here, $N$ represents the total number of sites in the chain. We note that $\alpha_l$ and $\alpha_l^\dagger$ are not, in the conventional sense, ladder operators, as their commutation relation is given by

$$[\alpha_l, \alpha_{l'}^\dagger] = \frac{1}{2m\omega_l N}\delta_{l,l'}.$$

We can then define ladder operators for the phonon modes as $m_l \equiv \sqrt{2m\omega_l N}\alpha_l$, which satisfy the commutation relation:

$$[m_l, m_{l'}^\dagger] = \delta_{l,l'}.$$

To demonstrate the equivalence between Eqn. (5) and the phonon part of Eqn. (1), we can plug the expressions for $\mathcal{U}$ and $\mathcal{P}$ into Hamiltonian Eqn. (5):

$$\begin{aligned}H &= \sum_n \frac{\mathcal{P}_n^2}{2m} + \sum_{j=1}^\nu \sum_n \frac{\lambda_1}{2}(\mathcal{U}_{n+j} - \mathcal{U}_n)^2 \\ &= \frac{1}{2}\sum_{l \neq 0}\omega_l(m_l^\dagger m_l + m_l m_l^\dagger) \\ &= \sum_{l \neq 0}\left[\omega_l m_l^\dagger m_l + \frac{1}{2}\omega_l\right].\end{aligned} \tag{11}$$

The second step involves noticing that $m_l m_l^\dagger = 1 + m_l^\dagger m_l$. We aim to have the annihilation operator at the end since it simplifies the calculation. This rearrangement procedure, using commutation relations to place all annihilation operators at the end, is known as normal ordering. It naturally introduces a constant as the zero-point energy, which we will ignore since it is subsumed into $\mathcal{V}_0$ and does not affect our calculations. Additionally, we note that the sum over normal modes $l$ is not infinite. We need to truncate it when the wavelength becomes short enough to approach the scale of atomic separation, in order to stay within the Brillouin zone. Hence, there are no concerns about ultraviolet divergence in this context. It is also not surprising that we obtain a sum of simple harmonic oscillators, as the potential term in Eqn. (5) is constructed to have a Hookean coupling.

### 2. Generalizing the simplest case into 2D and 3D lattices

We now consider the simplest case in 2D and 3D, where the system exhibits 1D couplings in all directions. In higher dimensions, we will have distortions in different directions denoted as $\mathcal{U}_\alpha(\mathbf{r})$. The simplified Hamiltonian in $d = 2, 3$ dimensions takes the following form:

$$H = \sum_n \frac{|\mathcal{P}_\mathbf{r}|^2}{2m} + \sum_\mathbf{r}\sum_{\alpha,\beta}\sum_{j=1}^\nu \frac{\lambda_j}{2}(\mathcal{U}_\alpha(\mathbf{r} + j\mathbf{e}_\beta) - \mathcal{U}_\alpha(\mathbf{r}))^2. \tag{12}$$

Moreover, we will assume that the solid is isotropic, which means that the sets of coupling constants $\lambda_j$ are the same in all directions. This assumption leads to the following frequency spectrum:

$$\omega_\alpha(\mathbf{k}) = \left( \sum_\beta \sum_{j=1}^\nu \frac{\lambda_j}{m} \left(2 - 2\cos k_\beta j a\right) \right)^{1/2}. \tag{13}$$

From this point onwards, we will focus solely on the simplified phonon Hamiltonian.

### C.  Construction of transmon Hamiltonian and interactions

The significant change we are making is in the model for the transmon qubit. Instead of treating it as a single harmonic oscillator, we will model it as a massless scalar field denoted by $\phi$:

$$\phi(t, \mathbf{x}) = \int \frac{d^d \mathbf{p}}{(2\pi)^{d/2} \sqrt{2\omega_\mathbf{p}}} \left( a_\mathbf{p} e^{-i(\omega_\mathbf{p} t - \mathbf{p}\cdot\mathbf{x})} + a_\mathbf{p}^\dagger e^{i(\omega_\mathbf{p} t - \mathbf{p}\cdot\mathbf{x})} \right). \tag{14}$$

Intuitively, the transmon qubit is piezoelectrically coupled to the resonator, which is described by the phonon modes. This allows us to drive the transmon qubit using electromagnetic waves, which in turn generate stress in the resonator and produce phonon modes (as discussed by Chu et al., 2017 [3]). In this context we model the electromagnetic field with only one polarization, as what was done in previous works, such as [3], which is naturally represented by a massless scalar field $\phi(x)$. With this simplification, the model we present below will be closest to the one considered in [4], a simple harmonic oscillator with some nonlinear interactions. The free Hamiltonian for a scalar field is well established in quantum field theory (as detailed by Coleman, 2018 [5]):

$$H_\phi = \int d^d \mathbf{x} \frac{1}{2} \left( \dot\phi(t, \mathbf{x})^2 + |\nabla\phi(t, \mathbf{x})|^2 \right). \tag{15}$$

We can now demonstrate that this Hamiltonian describes a "sum" of uncountably infinitely many harmonic oscillators.

$$\frac{1}{2}\int d^d\mathbf{x}\,\dot\phi(t,\mathbf{x})^2 = \frac{1}{2}\int \frac{d^d\mathbf{p}}{2\omega_\mathbf{p}}\left[ -\omega_\mathbf{p}^2 \left( a_\mathbf{p} a_{-\mathbf{p}} e^{-2i(\omega_\mathbf{p} t)} + a_\mathbf{p}^\dagger a_{-\mathbf{p}}^\dagger e^{2i(\omega_\mathbf{p} t)} \right) + \omega_\mathbf{p}^2 \left( a_\mathbf{p} a_\mathbf{p}^\dagger + a_\mathbf{p}^\dagger a_\mathbf{p} \right) \right], \tag{16}$$

$$\frac{1}{2}\int d^d\mathbf{x}\,|\nabla\phi(t,\mathbf{x})|^2 = \frac{1}{2}\int \frac{d^d\mathbf{p}}{2\omega_\mathbf{p}}|\mathbf{P}|^2\left[ \left( a_\mathbf{p} a_{-\mathbf{p}} e^{-2i(\omega_\mathbf{p} t)} + a_\mathbf{p}^\dagger a_{-\mathbf{p}}^\dagger e^{2i(\omega_\mathbf{p} t)} + a_\mathbf{p} a_\mathbf{p}^\dagger + a_\mathbf{p}^\dagger a_\mathbf{p} \right) \right]. \tag{17}$$

Plugging these expressions into the Hamiltonian, we can then proceed with our normal ordering procedure.

$$\begin{aligned} H_\phi &= \frac{1}{2}\int d^d\mathbf{p}\,\omega_\mathbf{p}(a_\mathbf{p}^\dagger a_\mathbf{p} + a_\mathbf{p} a_\mathbf{p}^\dagger) = \int d^d\mathbf{p}\,\omega_\mathbf{p}\left(a_\mathbf{p}^\dagger a_\mathbf{p} + \frac{1}{2}[a_\mathbf{p}, a_\mathbf{p}^\dagger]\right) \\ &= \int d^d\mathbf{p}\,\omega_\mathbf{p}\left(a_\mathbf{p}^\dagger a_\mathbf{p} + \frac{1}{2}\delta^{(d)}(0)\right) = \int d^d\mathbf{p}\,\omega_\mathbf{p}(a_\mathbf{p}^\dagger a_\mathbf{p}). \end{aligned} \tag{18}$$

Indeed, the presence of infinity in the normal ordering procedure is not a significant issue for several reasons. First, what can be measured in practice is the energy difference rather than the absolute energy. Second, we have considered the $\phi$ field or the transmon field within our solid, not in the vacuum of the universe. Therefore, this divergence, when placed in the context of the solid, will be transformed into a term proportional to the volume of our solid.

Contrary to this minor concern, the crucial point is that by treating the transmon as a scalar field, we reintroduce locality into the system. This locality is inherently embedded in the light-cone structure of $\phi$ as dictated by special relativity. Consequently, our new model of coupling the transmon qubit and the resonator through $\mathcal{U}$ and $\phi$ will always be constrained by locality. With these considerations in mind, we can now write down the full Hamiltonian.

$$\begin{aligned} H =& H_\mathcal{U} + H_\phi + H_I \\ =& \sum_\mathbf{r} \left( \frac{\mathcal{P}_\mathbf{r}^2}{2m} + \sum_j^\nu \sum_{\alpha,\beta} \frac{\lambda_j}{2} (\mathcal{U}_\alpha(\mathbf{r}) - \mathcal{U}_\alpha(\mathbf{r}+j\mathbf{e}_\beta))^2 \right) + \int d^d\mathbf{x}\frac{1}{2}\left(\dot\phi(t,\mathbf{x})^2 + |\nabla\phi(t,\mathbf{x})|^2\right) \\ &+ \int d^d\mathbf{x}\left[ \sum_\mathbf{r}\sum_\alpha \left( \mathcal{C}\phi(t,\mathbf{x})\mathcal{U}_\alpha(\mathbf{r}) \right) + \frac{h}{4!}\phi^4(t,\mathbf{x}) \right]. \end{aligned} \tag{19}$$

The $\phi^4$ interaction in the Hamiltonian is an analog of the Kerr nonlinearity term in Hamiltonian Eqn. (1). The coupling constant $h$ should be at the same order of magnitude as $\alpha$ there. Similarly, the quadratic coupling term $\mathcal{C}\phi(t,x)\mathcal{U}_n(t)$ is introduced to reproduce the couplings between the transmon qubit and the phonon modes, which were represented by terms such as $(q^\dagger m_k + \text{H.C.})$.

### D.  Lieb-Robinson bound for QRAM system

We will now study the Lieb–Robinson bound for Hamiltonian Eqn. (19). The Lieb–Robinson bound provides insights into the velocity of information propagation in a system governed by a specific Hamiltonian $H_{\Lambda_L}$. Here, $\Lambda_L$ represents the entire solid under consideration, with $L$ denoting the length scale of the solid. In the context of our construction, we will present a restatement of the Lieb–Robinson bound for harmonic lattice systems, as derived in [6]. To this end, we define the Weyl operators as follows:

$$W(f) \equiv \exp\left\{i \sum_{n \in \Lambda_L} \text{Re}[f(n)]q_n + \text{Im}[f(n)]p_n\right\}. \tag{20}$$

Furthermore, we denote the time evolution map of the Weyl operator as $\tau_t^{h,L}(W(f)) = e^{itH_L^h}W(f)e^{-itH_L^h}$, where $H_L^h$ represents the harmonic Hamiltonian defined as follows:

$$H_L^h = \sum_{n \in \Lambda_L} \frac{p_\mathbf{r}^2}{2m} + \frac{m\omega^2}{2}q_\mathbf{r}^2 + \sum_\alpha \frac{\lambda_\alpha}{2}(q_\mathbf{r} - q_{\mathbf{r}+\mathbf{e}_\alpha})^2. \tag{21}$$

In our case, we consider the special case where $\omega = 0$. This particular scenario is also discussed in detail in [6].

**Theorem I.1.** *For any $\mu > 0$, the estimate*

$$||[\tau_t^\Lambda(W(f)), W(g)]|| \leq C \sum_{x,y \in \Lambda_L} |f(x)||g(y)|e^{-\mu m\left[d(X,Y) - c_{\omega,\lambda}\max\left(\frac{2}{\mu}, e^{(\mu/2)+1}\right)|t|\right]}, \tag{22}$$

*holds for all functions $f, g \in l^2(\Lambda_L)$ and $t \in \mathbb{R}$. Here $\mu$ is a decay rate, and $d(x,y)$ is the distance between $x$ and $y$ calculated with the metric $d$. Moreover,*

$$C = \left(1 + c_{\omega,\lambda}e^{\mu/2} + c_{\omega,\lambda}^{-1}\right), \tag{23}$$

*with*

$$c_{\omega,\lambda} = \left(\sum_\alpha \frac{\lambda_\alpha}{m}\right)^{1/2}. \tag{24}$$

The detailed proof of the theorem can be found in [6]. For additional information, readers may also refer to [7]. Note that the commutator is exponentially suppressed if $d(X,Y) - c_{\omega,\lambda}\max\left(\frac{2}{\mu}, e^{(\mu/2)+1}\right)|t| > 0$. Given that $d(X,Y)$ represents the distance between $X$ and $Y$, this shows that the exponential suppression takes place when the velocity is greater than $c_{\omega,\lambda}\max\left(\frac{2}{\mu}, e^{(\mu/2)+1}\right)$. In quantum mechanics, if the commutator of two local operators approaches zero, it indicates that information does not propagate between them. We take the exponential suppression to be infinitely fast, such that there is little smearing outside the allowed regime. Since we are considering our system only at low temperature, the wave length of the sound wave $lambda$ should satisfy: $a \ll dx \ll \lambda$. This is explained in greater detail in later sections. Hence, if we take $a \sim 10^{-6}m$, then $\lambda$ should be around $1m \sim 10m$. Hence, if we take $d(X-Y) \sim \lambda$, then, over 1 wavelength, the commutator is going to be extremely suppressed. Now we consider the Lieb–Robinson bound for our Hamiltonian Eqn. (19).

**Proposition I.2.** For Hamiltonian Eqn. (12):

$$H_\mathcal{U} = \sum_\mathbf{r} \frac{|\mathcal{P}_\mathbf{r}|^2}{2m} + \sum_\mathbf{r}\sum_j^\nu\sum_{\alpha,\beta} \frac{\lambda_j}{2}(\mathcal{U}_\alpha(\mathbf{r}) - \mathcal{U}_\alpha(\mathbf{r}+j\mathbf{e}_\beta))^2, \tag{25}$$

we have the following Lieb–Robinson bound:

$$||[\tau_t^\Lambda(W(f)), W(g)]|| \leq C \sum_{x,y \in \Lambda_L} |f(x)||g(y)| e^{-\mu m \left[d(X,Y) - c_{\omega,\lambda} \max\left(\frac{2}{\mu}, e^{(\mu/2)+1}\right)|t|\right]}, \qquad (26)$$

where

$$c_{\omega,\lambda} = \left(d \sum_j^\nu \frac{\lambda_j}{m}\right)^{1/2}. \qquad (27)$$

*Proof.* Since the detailed proof of Theorem I.1 is given in [6] and the full proof here is almost the same as the proof there, we will not present the full proof. Instead, we will point out the change in some quantities calculated in the detailed proof. The good news is that the minor difference does not affect the calculation and results only in a minor change in numerical factors.

Because our Hamiltonian includes couplings between farther sites in all directions, the frequency spectrum of our system, as given in Eqn. (8) and Eqn. (13), differs from the frequency spectrum of the harmonic Hamiltonian in Eqn. (21). In [6], the authors denoted the frequency spectrum as $\gamma(k)$. As a result, for our case,

$$\gamma(k) = \left( \sum_\beta \sum_{j=1}^\nu \frac{\lambda_j}{m} (2 - 2\cos k_\beta j a) \right)^{1/2}. \qquad (28)$$

This means that in our case there are more terms summed over with different arguments in the cos function. However, changing the argument of cos does not affect its range of values. Therefore, the proof in [6] remains unchanged, except for the value of $c_{\omega,\lambda}$ (taking into account our isotropic assumption):

$$c_{\omega,\lambda} = \left(d \sum_j^\nu \frac{\lambda_j}{m}\right)^{1/2}. \qquad (29)$$

Moreover, the other terms in the Hamiltonian play a trivial role in the time evolution of $W(f)$. Since we are free to choose the function $f(x)$, for simplicity we can choose $f(x)$ to be a real-valued function. This simplifies the Weyl operator as follows:

$$W(f) = \exp\left\{ i \sum_{n \in \Lambda_L} f(n) q_n \right\}. \qquad (30)$$

Now, let us observe that the interaction Hamiltonian $H_I$ commutes with $W(f)$. Therefore, the time evolution operator generated by this term, denoted as $U_I$, acts trivially on $W(f)$: $e^{itH_I} W(f) e^{-itH_I} = W(f)$. Additionally, since $\phi$ and $\mathcal{U}$ commute, the Hamiltonian term $H_\phi$ also does not generate time evolution on $W(f)$. $\square$

Hence, the Lieb–Robinson velocity is

$$v_h(\mu) = \left(d \sum_{j=1}^\nu \lambda_j / m\right)^{1/2} \max\left[\frac{2}{\mu}, e^{(\mu/2)+1}\right]. \qquad (31)$$

We choose $1/2 \leq \mu \leq 1$, and by optimizing over $\mu$ we can determine the upper bound of Lieb–Robinson velocity as shown in [6]:

$$v_h \leq 4 \left(d \sum_{j=1}^\nu \frac{\lambda_j}{m}\right)^{1/2}. \qquad (32)$$

This is the speed limit for any information to travel in this QRAM system. As we can see, it depends on the sum of all coupling constants $\lambda_j$, which means that this is a completely intrinsic result for the solid. In other words, the couplings to the transmon

field do not affect the result at all. This should be clear when we proved that both $H_\phi$ and $H_I$ commutes with $W(f)$. The speed limit is then directly related to the upper bound $\nu$, which represents how we terminate the sum on $j$ in Eqn. (12).

Indeed, the Lieb–Robinson velocity provides a speed limit for the propagation of information in the QRAM system. It is determined by the sum of all coupling constants $\lambda_j$, which characterizes the properties of the solid. The couplings to the transmon field, represented by $H_\phi$ and $H_I$, do not affect the Lieb–Robinson velocity, as shown by the fact that they commute with $W(f)$. Therefore, the speed limit is intrinsic to the solid and is directly related to the choice of the upper bound $\nu$ in Eqn. (12), which determines the termination of the sum over $j$.

We write the speed limit for $2D$ and $3D$ in terms of our $1D$ results:

$$v_h^{(2d)} = \sqrt{2} v_h^{(1d)} = 4\sqrt{2}\bigg(\sum_{j=1}^{\nu} \frac{\lambda_j}{m}\bigg)^{1/2}; \qquad v_h^{(3d)} = \sqrt{3} v_h^{(1d)} = 4\sqrt{3}\bigg(\sum_{j=1}^{\nu} \frac{\lambda_j}{m}\bigg)^{1/2}. \tag{33}$$

These expressions demonstrate how the speed limits in higher dimensions are related to the 1D case and provide a scaling factor based on the dimensionality of the lattice.

## II. QUANTUM FIELD THEORY APPROACH TO ACHIEVE QUANTUM GATES

In order to establish the equivalence between the local Hamiltonian given by Eqn. (19) and the effective Hamiltonian described by Eqn. (1), it is necessary to coarse-grain the discrete Hamiltonian and obtain a full quantum field theory description. This involves treating the lattice as a continuum and expressing the fields and interactions in terms of continuous variables.

The process of coarse graining is typically achieved by taking a continuum limit, where the lattice spacing is taken to be infinitesimally small. This allows us to replace discrete sums with integrals over continuous variables and recover a field theory formulation.

Once we have obtained the quantum field theory description, we can study scattering processes represented by tree-level Feynman diagrams. These diagrams represent the scattering amplitudes and interactions between particles in the system. By analyzing these scattering processes, we can determine whether they correspond to the desired quantum gates generated by Eqn. (1). Additionally, by considering the scattering processes, we can calculate the clock cycle times.

Therefore, by coarse graining the discrete Hamiltonian, obtaining a quantum field theory description, and analyzing the scattering processes and clock cycle times, we can establish the equivalence between the local Hamiltonian and the effective Hamiltonian and show that the quantum gates generated by Eqn. (1) can also be generated by Eqn. (19).

### A.   Construction of a full quantum field theory

To analyze the QRAM system and determine clock cycle times for operations, we employ a coarse graining procedure. This leads to an effective Lagrangian density that describes the system as a general quartic theory involving the phonon and transmon fields.

It should be noted that the Lagrangian density presented here is not the only possible theory for constructing QRAM. Depending on the specific design and requirements, alternative Lagrangian densities can be formulated. For instance, incorporating classical communication systems operating at the speed of light would involve replacing the massless scalar field with the Maxwell field. With a basic understanding of electrodynamics, one can observe that signals propagate at the speed of light. This is a fundamental property of electromagnetic waves, as explained in standard textbooks on the subject such as [8–10].

Furthermore, in the context of quantum electrodynamics, one can put phonons in a $U(1)$ gauge field. This will introduce additional complexities due to minimal coupling which brings derivative interactions [5]. However, by evolving the field operator with the full Hamiltonian, one can observe that the derivative interactions arising from minimal coupling can be interpreted as a field translation operator. This leads to the addition of a massless vector field to the phonon field operator, which commutes with the phonon fields. Consequently, the calculation of the light cone structure should yield similar results, and the fundamental principles remain unchanged. If the photons are the carrier of the information, then there is not derivative interactions of $A_\mu$ involved. Hence, the $c = 3 \times 10^8$m/s is the ultimate speed limit. The key is to first make sure what's the medium that information travels in. And then based on the Lagrangian density of specific design, one can follow the same steps in this section to analyze the light cone structure. Our theory could also include classical field ingredients, where interactions could be generated through classical fields to model classical communication. In fact, if the Hamiltonian itself provides the speed of sound limit, using a classical communication tool could help improve the bound towards the light speed. This is consistent with the strategy used in [11].

The quartic theory we will present is a continuum generalization of the discrete model described earlier. It serves as a good example because it is a local theory that reproduces the same time scales calculated in [4]. Different models may require

different quantum field theories. However, the results obtained from this specific model are quite representative, and the light cone analysis can be applied to any specific model of interest. We will leave those detailed analysis of other models to future research.

*1.  Coarse graining*

Indeed, at low temperatures, where only modes with small frequencies are excited, the dominant behavior of the system is described by long-wavelength modes with wavelengths much larger than the lattice spacing ($\lambda \gg a$). In this regime, the motion of the atoms becomes collective, and the individual interactions among small particles become less relevant.

By considering the collective behavior of a group of atoms moving together, we can effectively average out the interactions among individual atoms and focus on the macroscopic behavior of the system. This is akin to taking a large $N$ limit, where $N$ represents the number of lattice sites or atoms, but applied in a different manner. Instead of directly considering the limit of a large number of lattice sites, we consider the behavior of averaged quantities over a coarse-grained distance $dx$, where $dx$ is much larger than the lattice spacing $a$.

Through this coarse-graining process, we construct a coarse-grained function $\mathcal{U}_\alpha(t, \mathbf{x})$ that describes the averaged distortions of the lattice at a larger scale. The choice of $\lambda \gg dx \gg a$ ensures that we capture the relevant collective behavior while effectively smoothing out the finer details at the scale of individual atoms.

This shift from discrete variables to a continuous function reflects a change in perspective achieved through coarse graining. It allows us to describe the system in terms of a continuous field theory, which is better suited for analyzing collective phenomena and understanding the emergent behavior of the system.

To construct the Lagrangian density for the phonon field, we start by assuming that there is a notion of locality. This implies that the functional $L_\mathcal{U}[\mathcal{U}\alpha(t, \mathbf{x})]$ can be written in the form of an integral over some density, denoted as $\int d^d \mathbf{x} \mathscr{L}\mathcal{U}$. Here, we consider the general case of $d$ dimensions, where we also assume a $d+1$ Lorentzian metric.

The Lagrangian density $\mathscr{L}\mathcal{U}$ depends on the distortion field $\mathcal{U}\alpha(t, \mathbf{x})$ and its derivatives. Specifically, it is a functional of the field itself $\mathcal{U}\alpha(t, \mathbf{x})$, its spatial derivatives, denoted by $\partial_\beta \mathcal{U}\alpha(t, \mathbf{x})$, and higher derivatives. The Lagrangian density may also depend on the time derivative $\dot{\mathcal{U}}_\alpha(t, \mathbf{x})$ of the field, capturing the dynamics of the system.

$$\mathscr{L}_\mathcal{U}\big[\mathcal{U}_\alpha(t, \mathbf{x}); \mathcal{U}'_\alpha(t, \mathbf{x}); \mathcal{U}''_\alpha(t, \mathbf{x}); \dot{\mathcal{U}}_\alpha(t, \mathbf{x}); \cdots\big] . \tag{34}$$

Indeed, the term $\left(\frac{\partial \mathcal{U}_\alpha}{\partial x^\beta}\right)^2$ in the Lagrangian density can be understood as the continuous version of the squared difference between neighboring lattice sites, $(\mathcal{U}_\alpha(\mathbf{r}) - \mathcal{U}_\alpha(\mathbf{r} + \mathbf{e}_\beta))^2$, in the discrete case. By including higher derivative terms, we effectively introduce couplings between farther lattice sites, allowing for a more accurate description of the system's behavior.

The concept of locality in this context means that even though we expand the Lagrangian density to include higher derivative terms, we can choose to truncate the expansion at a certain order. This truncation implies that we neglect terms involving even higher derivatives, which correspond to long-distance interactions. In other words, we assume that the most relevant interactions occur between neighboring lattice sites and that interactions beyond a certain distance are negligible. We will expand on this in a later section. So, the idea of locality is that while we make this expansion, at the end of the day we can terminate this expansion, such that we would not need to go to higher orders; hence, we do not have long-distance interactions.

Second, in the construction of the Lagrangian density $\mathscr{L}_\mathcal{U}$, we need to take into account the translational and rotational symmetries of the system. These symmetries impose constraints on the form of the Lagrangian density and ensure that it remains invariant under certain transformations.

The translational symmetry implies that the energy associated with a distortion field $\mathcal{U}\alpha(t, \mathbf{x})$ should not depend on the specific location in space. If we add a constant vector $\mathcal{C}\alpha$ to the distortion field at every point, the energy should remain unchanged. This translates into the requirement that the Lagrangian density $\mathscr{L}\mathcal{U}$ should not explicitly depend on the distortion field $\mathcal{U}\alpha(t, \mathbf{x})$ but rather on its derivatives or combinations thereof.

Similarly, the rotational symmetry implies that the energy should not change under rotations of the system. If we apply a rotation operator R to the distortion field $\mathcal{U}\alpha(t, \mathbf{x})$ (which corresponds to rotating the entire system), the energy should remain invariant. This imposes a constraint on the form of $\mathscr{L}\mathcal{U}$, ensuring that it has the same functional form when evaluated for the rotated distortion field $\mathrm{R}\left(\mathcal{U}_\alpha(t, \mathbf{x})\right)$.

$$\mathscr{L}_\mathcal{U}\big[\mathcal{U}_\alpha(t, \mathbf{x})\big] = \mathscr{L}_\mathcal{U}\big[\,\mathrm{R}\left(\mathcal{U}_\alpha(t, \mathbf{x})\right) + \mathcal{C}_\alpha\,\big] . \tag{35}$$

We now have the stability constraint. This ensures that the system is stable and that the energy is bounded from below. This constraint implies that there cannot be any linear terms in the Lagrangian density.

In the construction of the discrete Hamiltonian, the absence of linear terms is also explicitly required. This ensures that the expansion of the potential, as given in Eqn. (3), starts from quadratic terms and higher, with the linear terms vanishing.

Furthermore, in order for the system to be stable at the bottom of the potential, the quadratic potential term in the Lagrangian density should have the right sign. This ensures that the potential energy is minimized at the equilibrium position, corresponding to the stable configuration of the system.

Taking into account these stability considerations and the isotropic nature of the material, the most general form of the Lagrangian density, up to leading order of the derivative expansion in Fourier space with small wave vectors $k \to 0$, can be expressed as shown below.

$$\tilde{\mathscr{L}}_{\mathcal{U}} = \left[ \frac{\rho}{2} |\dot{\tilde{\mathcal{U}}}|^2 - \frac{\mu}{2} |\mathbf{k}|^2 |\tilde{\mathcal{U}}|^2 - \frac{\mu + \lambda}{2} |\mathbf{k} \cdot \tilde{\mathcal{U}}|^2 \right] . \tag{36}$$

We note that this construction is an approximation and is valid only within the regime of small wave vectors. Higher-order terms in the expansion, corresponding to shorter wavelengths, are neglected in this approach. Therefore, the resulting continuous theory is only accurate at the lowest order in $k$.

The two parameters in Eqn. (36) $\lambda$ and $\mu$ are called the Lamé coefficients, which are material-dependent parameters that characterize the elastic properties of the solid. The parameter $\rho$ represents the mass density of the material, where $m$ is the mass of an individual atom and $a$ is the lattice spacing. By Fourier transforming the Lagrangian density from momentum space to real space, we obtain the Lagrangian density below:

$$\mathscr{L}_{\mathcal{U}} = \frac{\rho}{2} |\dot{\mathcal{U}}|^2 - \sum_{\alpha,\beta} \left( \mu \mathcal{U}_{\alpha\beta} \mathcal{U}_{\alpha\beta} - \frac{\lambda}{2} \mathcal{U}_{\alpha\alpha} \mathcal{U}_{\beta\beta} \right) , \tag{37}$$

where $\mathcal{U}_{\alpha\beta} = \frac{1}{2} \left( \frac{\partial \mathcal{U}_\alpha}{\partial x^\beta} + \frac{\partial \mathcal{U}_\beta}{\partial x^\alpha} \right)$. We also applied the Einstein summation convention, hence in second and third terms, $\alpha$ and $\beta$ is summed over implicitly. Next, we will examine the Euler-Lagrange equation for the Lagrangian density:

$$\rho \frac{\partial^2 \mathcal{U}_\alpha}{\partial t^2} = (\mu + \lambda) \frac{\partial^2 \mathcal{U}_\beta}{\partial x^\alpha \partial x^\beta} + \mu \frac{\partial^2 \mathcal{U}_\alpha}{\partial x^\beta \partial x^\beta} . \tag{38}$$

The solutions to Eqn. (38) are polarized waves: $\mathcal{U}_\alpha(t, \mathbf{x}) = \epsilon_\alpha e^{i\mathbf{k} \cdot \mathbf{x} + \omega_\mathbf{k} t}$. Substituting the solutions back into Eqn. (38) yields the dispersion relation for each polarization. For the transverse polarization, $\epsilon \cdot \mathbf{k} = 0$, where $\epsilon$ represents the polarization vector and $\mathbf{k}$ is the wave vector, the dispersion relation is

$$\omega_\mathbf{k}^2 = \frac{\mu}{\rho} |\mathbf{k}|^2 . \tag{39}$$

In contrast, for the longitudinal polarization, $\mathbf{k} \sim \epsilon$, the dispersion relation is given by

$$\omega_\mathbf{k}^2 = \frac{2\mu + \lambda}{\rho} |\mathbf{k}|^2 . \tag{40}$$

Note that Eqn. (37) represents a general Lagrangian that includes "cross" couplings between different phonon modes. For simplicity, however, we will not use this general form and instead consider a simplified Lagrangian:

$$\mathscr{L}_{\mathcal{U}} = \frac{\rho}{2} |\dot{\mathcal{U}}|^2 - \sum_{\alpha,\beta}^{d} \frac{\lambda^{(d)}}{2} \left( \frac{\partial \mathcal{U}_\alpha}{\partial x^\beta} \right)^2 . \tag{41}$$

In the next section, we will demonstrate how Eqn. (41) is derived through a coarse-graining procedure from Eqn. (12). We will then analyze the resulting equation of motion for the phonon field.

### 2. *Equivalence between coupling constants in discrete construction and continuous construction*

Here we demonstrate the equivalence between the Lamé coefficients in the current context and the coupling constants $\lambda_j$ in Eqn. (12). We begin with the 1D case, where the Lagrangian density is given by the following expression:

$$\mathcal{L}_{\mathcal{U}}^{(1D)} = \frac{\rho}{2}\left(\frac{\partial \mathcal{U}}{\partial t}\right)^2 - \frac{\lambda^{(1d)}}{2}\left(\frac{\partial \mathcal{U}}{\partial x}\right)^2, \tag{42}$$

and the equation of motion can be expressed as follows:

$$\rho \frac{\partial^2 \mathcal{U}}{\partial t^2} = -\lambda^{(1d)} \frac{\partial^2 \mathcal{U}}{\partial x^2}. \tag{43}$$

It is evident that Eqn. (43) corresponds to the continuous version of the equation of motion in the discrete case given by Eqn. (5). The dispersion relation obtained from the equation of motion, Eqn. (43), is given by

$$\omega = \sqrt{\frac{\lambda^{(1d)}}{\rho}} k. \tag{44}$$

We can obtain this dispersion relation by expanding the frequency spectrum in Eqn. (8) for small values of $k$ (i.e., $k \to 0$). Specifically, we have the following expansion:

$$\omega = \left(\sum_{j=1}^{\nu} \frac{\lambda_j}{m}(2 - 2\cos k_l j a)\right)^{1/2} \approx \left(\sum_{j=1}^{\nu} \frac{\lambda_j}{m}\right)^{1/2} jak = \left(\frac{\sum_{j=1}^{\nu} \lambda_j a j^2}{\rho}\right)^{1/2} k = \sqrt{\frac{\lambda^{(1d)}}{\rho}} k. \tag{45}$$

Therefore, we can establish the equivalence between the Lamé coefficient $\mu$ in the continuous case and the coupling constants $\lambda_j$ in the discrete case for the 1D scenario:

$$\lambda^{(1d)} = \sum_{j=1}^{\nu} \left(\lambda_j a j^2\right). \tag{46}$$

We can extend this generalization to the $d$-dimensional case (where we consider $d = 2$ or $d = 3$). The coarse-grained Lagrangian obtained from Eqn. (12) is given by Eqn. (41). This simplified model includes only transverse waves, and the corresponding equation of motion can be expressed as

$$\rho \frac{\partial^2 \mathcal{U}_\alpha}{\partial t^2} = \lambda^{(d)} \frac{\partial^2 \mathcal{U}_\alpha}{\partial x^\beta \partial x^\beta}. \tag{47}$$

The equation of motion yields the following dispersion relation for the transverse waves ($k = |\mathbf{k}|$):

$$\omega = \sqrt{\frac{\lambda^{(d)}}{\rho}} k = \sqrt{\frac{d\lambda^{(1d)}}{\rho}} k. \tag{48}$$

Expanding the frequency spectrum Eqn. (13) for small wave numbers $k = |\mathbf{k}|$, we have

$$\omega(k) = \left(\sum_\beta \sum_{j=1}^{\nu} \frac{\lambda_j}{m}(2 - 2\cos k_\beta j a)\right)^{1/2} \approx \sqrt{d}\left(\sum_{j=1}^{\nu} \frac{j^2 \lambda_j}{m}\right)^{1/2} ak. \tag{49}$$

So, in $d$-dimensions, we have the following equivalence between the Lamé coefficients $\lambda$ and the coupling constants $\lambda_j$:

$$\lambda^{(d)} = d\sum_{j=1}^{\nu} \lambda_j a j^2. \tag{50}$$

### 3. Canonical quantization of phonon fields

The next step involves the canonical quantization of the system, which allows us to obtain the phonon field operator. This operator enables the creation and annihilation of phonons, with the index $s$ denoting the polarization of the phonons.

$$\mathcal{U}_\alpha(x) = \int \frac{d^d\mathbf{k}}{(2\pi)^{d/2}\sqrt{2\rho\omega_s(\mathbf{k})}} \epsilon^s_\alpha \left( m_s(\mathbf{k})e^{-i\mathbf{k}\cdot\mathbf{x}-i\omega_s(\mathbf{k})t} + m^\dagger_s(\mathbf{k})e^{i\mathbf{k}\cdot\mathbf{x}+i\omega_s(\mathbf{k})t} \right). \tag{51}$$

The canonical momentum, denoted as $\pi_\alpha(x)$, is given by $\pi_\alpha(x) = \rho\dot{\mathcal{U}}\alpha(x)$, where, again, $\rho$ is the mass density and $\dot{\mathcal{U}}\alpha(x)$ represents the time derivative of the distortion field $\mathcal{U}_\alpha(x)$.

$$\mathcal{P}_\alpha(x) = \int \frac{d^d\mathbf{k}}{(2\pi)^{d/2}\sqrt{2\rho\omega_s(\mathbf{k})}} \epsilon^s_\alpha \left( -i\rho\omega_s(\mathbf{k})m_s(\mathbf{k})e^{-i\mathbf{k}\cdot\mathbf{x}-i\omega_s(\mathbf{k})t} + i\rho\omega_s(\mathbf{k})m^\dagger_s(\mathbf{k})e^{i\mathbf{k}\cdot\mathbf{x}+i\omega_s(\mathbf{k})t} \right). \tag{52}$$

The canonical quantization procedure imposes the canonical commutation relation, which is given by

$$[m_s(\mathbf{k}), m^\dagger_{s'}(\mathbf{k}')] = \delta_{s,s'}\delta^{(d)}(\mathbf{k}-\mathbf{k}') \qquad \text{and} \qquad [m_s(\mathbf{k}), m_{s'}(\mathbf{k}')] = [m^\dagger_s(\mathbf{k}), m^\dagger_{s'}(\mathbf{k}')] = 0. \tag{53}$$

The equal-time commutation relations can be calculated as follows:

$$[\mathcal{U}_\alpha(t,\mathbf{x}), \mathcal{U}_\beta(t,\mathbf{y})] = \int \frac{d^d\mathbf{k}}{(2\pi)^d (2\rho\omega_s(\mathbf{k}))} \delta_{\alpha,\beta} \left( e^{-i\mathbf{k}\cdot(\mathbf{x}-\mathbf{y})} - e^{i\mathbf{k}\cdot(\mathbf{x}-\mathbf{y})} \right) = 0. \tag{54}$$

The obtained result is due to the fact that the integrand in the calculation is an odd function. Also,

$$\begin{aligned}
[\pi_\alpha(t,\mathbf{x}), \mathcal{U}_\beta(t,\mathbf{y})] &= \frac{\partial}{\partial x^0}[\mathcal{U}_\alpha(t,\mathbf{x}), \mathcal{U}_\beta(t,\mathbf{y})] \\
&= \delta_{\alpha,\beta} \int \frac{d^d\mathbf{k}}{(2\pi)^d (2\rho\omega_s(\mathbf{k}))} \left( -i\rho\omega_s(\mathbf{k})e^{-i\mathbf{k}\cdot(\mathbf{x}-\mathbf{y})} - i\rho\omega_s(\mathbf{k})e^{i\mathbf{k}\cdot(\mathbf{x}-\mathbf{y})} \right) \\
&= -i\delta_{\alpha,\beta} \int \frac{d^d\mathbf{k}}{(2\pi)^d (2)} \left( e^{-i\mathbf{k}\cdot(\mathbf{x}-\mathbf{y})} + e^{i\mathbf{k}\cdot(\mathbf{x}-\mathbf{y})} \right) \\
&= -i\delta_{\alpha,\beta} \int \frac{d^d\mathbf{k}}{(2\pi)^d} \left( e^{-i\mathbf{k}\cdot(\mathbf{x}-\mathbf{y})} \right) = -i\delta_{\alpha,\beta}\delta^{(d)}(\mathbf{x}-\mathbf{y}),
\end{aligned} \tag{55}$$

where the fourth step is from the fact that $e^{-i\mathbf{k}\cdot(\mathbf{x}-\mathbf{y})} + e^{i\mathbf{k}\cdot(\mathbf{x}-\mathbf{y})}$ is even.

$$\begin{aligned}
[\pi_\alpha(t,\mathbf{x}), \pi_\beta(t,\mathbf{y})] &= \delta_{\alpha,\beta} \int \frac{d^d\mathbf{k}}{(2\pi)^d (2\rho\omega_s(\mathbf{k}))} \left( -(\rho\omega_s(\mathbf{k}))^2 e^{-i\mathbf{k}\cdot(\mathbf{x}-\mathbf{y})} + (\rho\omega_s(\mathbf{k}))^2 e^{i\mathbf{k}\cdot(\mathbf{x}-\mathbf{y})} \right) \\
&= \delta_{\alpha,\beta} \int \frac{d^d\mathbf{k}}{(2\pi)^d (2)} \left( -\rho\omega_s(\mathbf{k})e^{-i\mathbf{k}\cdot(\mathbf{x}-\mathbf{y})} + \rho\omega_s(\mathbf{k})e^{i\mathbf{k}\cdot(\mathbf{x}-\mathbf{y})} \right) = 0.
\end{aligned} \tag{56}$$

The Hamiltonian density is obtained by a Legendre transformation: $\mathscr{H}_\mathcal{U} = \pi_\alpha\dot{\mathcal{U}}_\alpha - \mathscr{L}_\mathcal{U}$:

$$\mathscr{H}_\mathcal{U} = \frac{\rho}{2}|\dot{\mathcal{U}}|^2 + \frac{\mu}{2}\frac{\partial\mathcal{U}_\alpha}{\partial x^\beta}\frac{\partial\mathcal{U}_\alpha}{\partial x^\beta} + \frac{\lambda}{2}\frac{\partial\mathcal{U}_\alpha}{\partial x^\alpha}\frac{\partial\mathcal{U}_\beta}{\partial x^\beta}. \tag{57}$$

By substituting Eqn. (51) into Eqn. (57), performing normal ordering, and neglecting the zero-point energy, we obtain the following Hamiltonian, which represents an integration over infinitely many harmonic oscillators due to the coarse-graining procedure:

$$H = \sum_s \int d^d\mathbf{k}\,\omega_s(\mathbf{k}) m_s^\dagger(\mathbf{k}) m_s(\mathbf{k}) \ . \tag{58}$$

Therefore, the full Lagrangian density, considering phonon fields, transmon field, and their interaction, is given by

$$\mathscr{L} = \frac{\rho}{2}|\dot{\mathcal{U}}|^2 - \frac{\mu}{2}\frac{\partial \mathcal{U}_\alpha}{\partial x^\beta}\frac{\partial \mathcal{U}_\alpha}{\partial x^\beta} - \frac{\lambda}{2}\frac{\partial \mathcal{U}_\alpha}{\partial x^\alpha}\frac{\partial \mathcal{U}_\beta}{\partial x^\beta} + \frac{1}{2}\partial_\mu\phi\partial^\mu\phi - \sum_\alpha \mathcal{C}\phi\mathcal{U}_\alpha - \frac{h}{4!}\phi^4 \ . \tag{59}$$

On the contrary, the Lagrangian that is obtained by coarse-graining the simplified Hamiltonian Eqn. (19) is given by

$$\mathscr{L} = \frac{\rho}{2}|\dot{\mathcal{U}}|^2 - \sum_{\alpha,\beta}^d \frac{\lambda^{(d)}}{2}\left(\frac{\partial \mathcal{U}_\alpha}{\partial x^\beta}\right)^2 + \frac{1}{2}\partial_\mu\phi\partial^\mu\phi - \sum_\alpha \mathcal{C}\phi\mathcal{U}_\alpha - \frac{h}{4!}\phi^4 \ . \tag{60}$$

### B. Locality through quantum field theory

The quantum field theory formulation of this system also imposes an explicit locality constraint similar to the Lieb–Robinson bound. Furthermore, we can establish the equivalence between the Lieb–Robinson bound and our locality bound through the process of coarse graining. Note that the following analysis is based on the general Hamiltonian given by Eqn. (19). Later, we can easily extrapolate these results to our simplified Hamiltonian given by Eqn. (60).

We will examine the commutator $[\pi_\alpha(t_x, \mathbf{x}), \mathcal{U}_\beta(t_y, \mathbf{y})]$ where we are interested in the more general case of nonequal times, unlike the equal time commutator in Eqn. (55). This corresponds to the concept of the $\tau$ map in the Lieb–Robinson bound. In quantum field theory, all operators are in the Heisenberg picture, so considering an operator at $t_y$ can be seen as evolved into the future, similar to the $\tau$ map in the Lieb–Robinson bound calculation. We note that we will perform this calculation for the general field theory case and then specialize to the simplified case later on.

To verify that information cannot travel faster than the speed of light, we examine this commutator. According to special relativity, if $x$ and $y$ have a spacelike separation, the commutator must be zero. The commutator can be expressed as

$$[\pi_\alpha(t_x, \mathbf{x}), \mathcal{U}_\beta(t_y, \mathbf{y})] = \frac{\partial}{\partial x^0}[\mathcal{U}_\alpha(t_x, \mathbf{x}), \mathcal{U}_\beta(t_y, \mathbf{y})].$$

Therefore, it suffices to show that $[\mathcal{U}_\alpha(t_x, \mathbf{x}), \mathcal{U}_\beta(t_y, \mathbf{y})] = 0$. For simplicity, we define $\mathcal{U}_\alpha(x) = \mathcal{U}\alpha^+(x) + \mathcal{U}_\alpha^-(x)$, where $\mathcal{U}_\alpha^+(x)$ and $\mathcal{U}_\alpha^-(x)$ are given by

$$\mathcal{U}_\alpha^-(x) = \int \frac{d^d\mathbf{k}}{(2\pi)^{d/2}\sqrt{2\rho\omega_s(\mathbf{k})}} \epsilon_\alpha^s \left( m_s(\mathbf{k}) e^{-i\mathbf{k}\cdot\mathbf{x} - i\omega_s(\mathbf{k})t} \right), \tag{61}$$

and

$$\mathcal{U}_\alpha^+(x) = \int \frac{d^d\mathbf{k}}{(2\pi)^{d/2}\sqrt{2\rho\omega_s(\mathbf{k})}} \epsilon_\alpha^s \left( m_s^\dagger(\mathbf{k}) e^{i\mathbf{k}\cdot\mathbf{x} + i\omega_s(\mathbf{k})t} \right). \tag{62}$$

The commutator of $\mathcal{U}\alpha^-(x)$ and $\mathcal{U}\alpha^+(x)$ can be calculated as follows:

$$\begin{aligned}\left[\mathcal{U}_\alpha^-(t_x, \mathbf{x}), \mathcal{U}_\beta^+(t_y, \mathbf{y})\right] &= \delta_{\alpha,\beta} \int \frac{d^d\mathbf{k}}{(2\pi)^{d/2}\sqrt{2\rho\omega_s(\mathbf{k})}} \int \frac{d^d\mathbf{k}'}{(2\pi)^{d/2}\sqrt{2\rho\omega_s(\mathbf{k}')}} e^{-ik\cdot x} e^{ik'\cdot y}\left[m_s(\mathbf{k}), m_s^\dagger(\mathbf{k}')\right] \\ &= \delta_{\alpha,\beta} \int \frac{d^d\mathbf{k}}{(2\pi)^{d/2}\sqrt{2\rho\omega_s(\mathbf{k})}} \int \frac{d^d\mathbf{k}'}{(2\pi)^{d/2}\sqrt{2\rho\omega_s(\mathbf{k}')}} e^{-ik\cdot x} e^{ik'\cdot y} \delta^{(d)}(\mathbf{k} - \mathbf{k}') \\ &= \delta_{\alpha,\beta} \int \frac{d^d\mathbf{k}}{(2\pi)^d 2\omega_s(\mathbf{k})} e^{-ik\cdot(x-y)} \equiv \Delta_+(x - y) \ .\end{aligned} \tag{63}$$

Expressing $[\mathcal{U}\alpha(t_x, \mathbf{x}), \mathcal{U}\beta(t_y, \mathbf{y})]$ in terms of the function $\Delta_+(x - y)$ provides a convenient form for further analysis:

$$\begin{aligned}\left[\mathcal{U}_\alpha(t_x,\mathbf{x}),\mathcal{U}_\beta(t_y,\mathbf{y})\right] &= \left[\mathcal{U}_\alpha^+(x)+\mathcal{U}_\alpha^-(x),\mathcal{U}_\beta^+(y)+\mathcal{U}_\beta^-(y)\right] \\ &= \left[\mathcal{U}_\alpha^-(t_x,\mathbf{x}),\mathcal{U}_\beta^+(t_y,\mathbf{y})\right]+\left[\mathcal{U}_\alpha^+(t_x,\mathbf{x}),\mathcal{U}_\beta^-(t_y,\mathbf{y})\right] \\ &= \Delta_+(x-y)-\Delta_+(y-x)\equiv i\Delta(x-y)\,.\end{aligned} \quad (64)$$

An important observation is that both $\Delta_+(x-y)$ and $i\Delta(x-y)$ are Lorentz scalar functions. This is evident from the fact that the argument of the exponential is a Lorentz scalar and the integral measure is constructed to be Lorentz invariant. Another useful fact is that if $(x-y)^2<0$, then $(y-x)=\Lambda(x-y)$. In other words, spacelike vectors can be transformed to their negatives by using a Lorentz transformation. This behavior arises from the properties of irreducible representations of the Lorentz group. For non-spacelike vectors, inverting the vector corresponds to changing the energy of the state, leading to a different irreducible representation. For spacelike vectors, however, inverting the vector still keeps it within the same irreducible representation. For further details on this topic, refer to [5] and [12]. With this fact, we can rewrite Eqn. (64):

$$\begin{aligned}\left[\mathcal{U}_\alpha(t_x,\mathbf{x}),\mathcal{U}_\beta(t_y,\mathbf{y})\right] &= \Delta_+(x-y)-\Delta_+(y-x)=\Delta_+(x-y)-\Delta_+(\Lambda(x-y)) \\ &= \Delta_+(x-y)-\Delta_+(x-y)=i\Delta(x-y)=0\,, \qquad \text{if } (x-y)^2<0\,.\end{aligned} \quad (65)$$

Eqn. (65) leads us to the desired result:

$$\left[\pi_\alpha(t_x,\mathbf{x}),\mathcal{U}_\beta(t_y,\mathbf{y})\right]=i\frac{\partial}{\partial x^0}\Delta(x-y)=0\,,\qquad \text{if } (x-y)^2<0\,. \quad (66)$$

Eqn. (66) explicitly demonstrates that there is no possibility of information propagating faster than the speed of light, and our construction remains consistent with special relativity. However, this is a very loose limit. In Eqn. (31) the Lieb–Robinson velocity depends only on the coupling constants within the solid itself. If we consider it to be on the order of the speed of sound in the solid, this limit would be much lower than the speed of light. Therefore, we need to examine $\left[\pi_\alpha(t_x,\mathbf{x}),\mathcal{U}_\beta(t_y,\mathbf{y})\right]$ more closely when $(x-y)^2>0$:

$$\begin{aligned}\left[\pi_\alpha(x),\mathcal{U}_\beta(y)\right] &= \frac{\partial}{\partial x^0}\left[\mathcal{U}_\alpha(x),\mathcal{U}_\beta(y)\right] \\ &= \delta_{\alpha,\beta}\int\frac{d^d\mathbf{k}}{(2\pi)^d(2\rho\omega_s(\mathbf{k}))}\left(-i\rho\omega_s(\mathbf{k})e^{-ik\cdot(x-y)}-i\rho\omega_s(\mathbf{k})e^{ik\cdot(x-y)}\right) \\ &= -i\delta_{\alpha,\beta}\int\frac{d^d\mathbf{k}}{(2\pi)^d(2)}\left(e^{-ik\cdot(x-y)}+e^{ik\cdot(x-y)}\right) \\ &= -i\delta_{\alpha,\beta}\int\frac{d^d\mathbf{k}}{(2\pi)^d}\left(e^{-ik\cdot(x-y)}\right)\,.\end{aligned} \quad (67)$$

This integral is explicitly evaluated for $d=3$ in [5]. We would like to perform the calculation for the cases of $d=1,2,3$. To begin, we transform to polar coordinates. Let us define $r=|\mathbf{x}-\mathbf{y}|$, $k=|\mathbf{k}|$, and $t=t_x-t_y$. Additionally, we recall the dispersion relations for transverse and longitudinal polarizations: $\omega_t^2(k)=\frac{\mu}{\rho}k^2$ and $\omega_l^2(k)=\frac{2\mu+\lambda}{\rho}k^2$. For the 3D case, the integral takes the following form:

$$\begin{aligned}\int\frac{d^3\mathbf{k}}{(2\pi)^3}e^{i\mathbf{k}\cdot(\mathbf{x}-\mathbf{y})}e^{-i\omega_s(\mathbf{k})(t_x-t_y)} &= \int_0^\infty\frac{k^2dk}{(2\pi)^3}e^{-i\omega_s(\mathbf{k})t}\int_0^\pi e^{ikr\cos\theta}\sin\theta d\theta\int_0^{2\pi}d\varphi \\ &= \int_0^\infty\frac{dk}{(2\pi)^2}ke^{-i\omega_s(\mathbf{k})t}\frac{e^{ikr}-e^{-ikr}}{ir}=-\frac{i}{2\pi^2 r}\int_{-\infty}^\infty dk\, k e^{ikr-i\omega_s(\mathbf{k})t}\,.\end{aligned} \quad (68)$$

In the case of 2D, the integral over $\theta$ yields a first-class Bessel function: $\int_0^{2\pi}e^{ikr\cos\theta}d\theta=J_0(kr)$. Since we are working in the long wavelength limit, we can expand the Bessel function in a series for $k\to 0$ as follows:

$$J_0(z)=1-\frac{z^2}{4}+\frac{z^4}{64}+\mathcal{O}(z^6).$$

Approximating this series with a simpler function, we can use the expression

$$\frac{e^{ikr} - e^{-ikr}}{ikr} = 2 - \frac{k^2 r^2}{3} + \frac{k^4 r^4}{60}.$$

The only difference between the two series is the numerical factors, and these differences are not significant. The reason is that all factors at the same level of expansion are of the same order. Therefore, we can approximate the Bessel function with $\frac{e^{ikr} - e^{-ikr}}{ikr}$:

$$\int \frac{d^2 \mathbf{k}}{(2\pi)^2} e^{i\mathbf{k}\cdot(\mathbf{x}-\mathbf{y})} e^{-i\omega_s(\mathbf{k})(t_x - t_y)} = \int_0^\infty \frac{k dk}{(2\pi)^2} e^{-i\omega_s(\mathbf{k})t} \int_0^{2\pi} e^{ikr \cos\theta} d\theta \qquad (69)$$
$$\approx \int_0^\infty \frac{dk}{(2\pi)^2} e^{-i\omega_s(\mathbf{k})t} \frac{e^{ikr} - e^{-ikr}}{ir} = -\frac{i}{2\pi^2 r} \int_{-\infty}^\infty dk\, e^{ikr - i\omega_s(\mathbf{k})t}.$$

In the case of 1D, the integral simplifies as follows:

$$\int \frac{dk}{2\pi} e^{ik(x-y)} e^{-i\omega_k(t_x - t_y)} = \int_{-\infty}^\infty \frac{dk}{\pi} e^{ikr - i\omega_k t}. \qquad (70)$$

Note that the integrals for the $d = 1$ and $d = 2$ cases are the same. Next, we perform an analytical continuation of the three integrals to the complex plane by substituting $k$ with the complex variable $z = x + iy$. However, we note that the integrand is not analytic in $k$ because of the dispersion relation $\omega_s(k) \propto k$ for all polarizations $s$. This leads to a branch point at $k = 0$. To avoid encountering issues with the branch point, we introduce a small quantity $\epsilon$ such that the branch points are shifted to $k = \pm i\epsilon$. Consequently, the frequency will take different values on the two sides of the branch cut.

$$\omega_t(k) = \begin{cases} i\sqrt{\mu/\rho}\sqrt{y^2 - \epsilon^2}, & x = 0+ \\ -i\sqrt{\mu/\rho}\sqrt{y^2 - \epsilon^2}, & x = 0- \end{cases}$$
$$\omega_l(k) = \begin{cases} i\sqrt{(2\mu + \lambda)/\rho}\sqrt{y^2 - \epsilon^2}, & x = 0+ \\ -i\sqrt{(2\mu + \lambda)/\rho}\sqrt{y^2 - \epsilon^2}, & x = 0- \end{cases} \qquad (71)$$

The contour, which accounts for the analytical behavior of the function, is depicted in FIG. 1. The $x$-axis is naturally a part of the integral contour $C$. To complete the contour, we include a large semicircle located above the $x$-axis, slightly distorted to circumvent the branch cut and ensure the proper analytic continuation of the function.

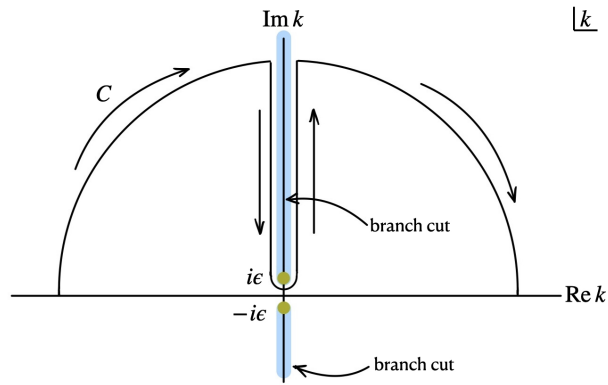

FIG. 1. Contour with $x$-axis and distorted large semicircle above the $x$-axis

The integration along the full $x$-axis can be expressed as the sum of the contributions from the remaining parts of the contour. Let us consider the parameterization of $k$ using $R\cos\theta + iR\sin\theta$ for the two large arcs. The left arc corresponds to $\pi \geq \theta \geq \pi/2$,

while the right arc corresponds to $\pi/2 \geq \theta \geq 0$. However, as we take the limit $R \to \infty$, an exponential suppression $e^{-Rr\sin\theta}$ arises, rendering the contributions from these arcs negligible. Similarly, the small semicircle does not contribute either.

Let us consider the case of transverse polarization, where we replace $\sqrt{\mu/\rho}$ with $\sqrt{(2\mu+\lambda)/\rho}$ for the longitudinal wave. What remains in the contour integrals (for different dimensions) are the following expressions:

$$\int_{-\infty}^{\infty} dk k e^{ikr-i\omega_s(\mathbf{k})t} = \left[\int_{\infty}^{\epsilon} dy y e^{-ry-\sqrt{\mu/\rho}\sqrt{y^2-\epsilon^2}t'} + \int_{\epsilon}^{\infty} dy y e^{-ry+\sqrt{\mu/\rho}\sqrt{y^2-\epsilon^2}t'}\right]$$
$$= \int_{\epsilon}^{\infty} dy y e^{-ry} \sinh\left(\sqrt{\mu/\rho}\sqrt{y^2-\epsilon^2}t'\right), \tag{72}$$

$$\int_{-\infty}^{\infty} dk e^{ikr-i\omega_k t} = \left[\int_{\infty}^{\epsilon} dy e^{-ry-\sqrt{\mu/\rho}\sqrt{y^2-\epsilon^2}t'} + \int_{\epsilon}^{\infty} dy e^{-ry+\sqrt{\mu/\rho}\sqrt{y^2-\epsilon^2}t'}\right]$$
$$= \int_{\epsilon}^{\infty} dy e^{-ry} \sinh\left(\sqrt{\mu/\rho}\sqrt{y^2-\epsilon^2}t'\right). \tag{73}$$

These integrals are not doable, but we can obtain upper bounds by considering only the increasing part of the sinh function and replacing $\sqrt{y^2-\epsilon^2}$ with $y$. This overestimates the integrals, and we obtain the following bounds for 1D, 2D, and 3D dimensions:

$$[\mathcal{U}_\beta(y), \pi_\alpha(x)] < \frac{\delta_{\alpha,\beta}}{2\pi^2 r} \int_\epsilon^\infty dy y e^{-(r-\sqrt{\mu/\rho}t)y}$$
$$< \frac{\delta_{\alpha,\beta}}{2\pi^2 r} e^{-(r-\sqrt{\mu/\rho}t)\epsilon} \left(\frac{1}{(r-\sqrt{\mu/\rho}t)^2} + \frac{\epsilon}{(r-\sqrt{\mu/\rho}t)}\right), \quad \text{(for 3 dimension)}. \tag{74}$$

$$[\mathcal{U}_\beta(y), \pi_\alpha(x)] < \frac{\delta_{\alpha,\beta}}{2\pi^2 r} \int_\epsilon^\infty dy e^{-(r-\sqrt{\mu/\rho}t)y} = \frac{\delta_{\alpha,\beta}}{2\pi^2 r} \frac{e^{-(r-\sqrt{\mu/\rho}t)\epsilon}}{(r-\sqrt{\mu/\rho}t)}, \quad \text{(for 2 dimension)}. \tag{75}$$

$$[\mathcal{U}_\beta(y), \pi_\alpha(x)] < \frac{1}{\pi} \int_\epsilon^\infty dy e^{-(r-\sqrt{\mu/\rho}t)y} = \frac{1}{\pi} \frac{e^{-(r-\sqrt{\mu/\rho}t)\epsilon}}{(r-\sqrt{\mu/\rho}t)}, \quad \text{(for 1 dimension)}. \tag{76}$$

Before taking the limit on $\epsilon$, we observe that there is exponential suppression if $(r-\sqrt{\mu/\rho}t) > 0$, similar to the behavior described by the Lieb-Robinson bound. This consistency reinforces the idea that the speed limit of information propagation is an intrinsic property of the solid itself. The Lamé coefficients in Eqn. (92) represent properties of the solid, while the coupling constants $\lambda_j$ are determined by the internal structure of the resonator. In other words, even in the full quantum field theory formulation where the solid is coupled to the $\phi$ (transmon) field, the speed of information propagation remains unchanged. It can be verified that the interaction term $\mathcal{L}_I$ in Eqn. (59) commutes with the field $\mathcal{U}_\alpha(x)$ and the free field part of $\phi$ is irrelevant to $\mathcal{U}_\alpha(x)$. For our simplified model, we can derive the following results.

$$[\mathcal{U}_\beta(y), \pi_\alpha(x)] < \frac{\delta_{\alpha,\beta}}{2\pi^2 r} e^{-(r-\sqrt{\lambda^{(3d)}/\rho}t)\epsilon} \left(\frac{1}{(r-\sqrt{\lambda^{(3d)}/\rho}t)^2} + \frac{\epsilon}{(r-\sqrt{\lambda^{(3d)}/\rho}t)}\right), \quad \text{(for 3 dimension)}. \tag{77}$$

$$[\mathcal{U}_\beta(y), \pi_\alpha(x)] < \frac{\delta_{\alpha,\beta}}{2\pi^2 r} \frac{e^{-(r-\sqrt{\lambda^{(2d)}/\rho}t)\epsilon}}{(r-\sqrt{\lambda^{(2d)}/\rho}t)}, \quad \text{(for 2 dimension)}. \tag{78}$$

$$[\mathcal{U}_\beta(y), \pi_\alpha(x)] < \frac{1}{\pi} \frac{e^{-(r-\sqrt{\lambda^{(1d)}/\rho}t)\epsilon}}{(r-\sqrt{\lambda^{(1d)}/\rho}t)}, \quad \text{(for 1 dimension)}. \tag{79}$$

Therefore, the speed limits derived in this analysis bear resemblance to the results obtained in the calculation of the Lieb–Robinson bound, as presented in Eqn. (32). Indeed, by applying the equivalence relation between the Lamé coefficients in the

continuous theory and the coupling constants $\lambda_j$ in the discrete theory, and focusing on the case where $\nu = 1$, we observe that the two velocity bounds are identical. However, discrepancies arise when $j$ can take values greater than 1. This is due to the fact that in the discrete theory, when considering couplings between more distant sites, higher-order derivatives in the field theory analysis need to be taken into account in order to ensure accuracy.

In addition, the relations between velocity limits in different dimensions remain the same as in the discrete case:

$$v^{(d)} = \left(\frac{\lambda^{(d)}}{\rho}\right)^{1/2} = \sqrt{d}\, v^{(1d)} = \sqrt{d}\left(\frac{\lambda^{(1d)}}{\rho}\right)^{1/2}, \quad \text{for } d = 1, 2, 3. \tag{80}$$

### C. Feynman diagrams and quantum gates

In this section we demonstrate that the Feynman diagrams of Eqn. (60) correspond to the quantum gates generated by Eqn. (1). The Feynman rules of Eqn. (60) involve two vertices, as illustrated in FIG. 2.

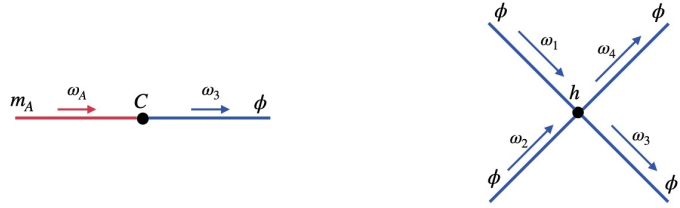

FIG. 2. The vertices in the Feynman rules of the Lagrangian density Eqn. (60) have certain energy conservation requirements. For the first diagram, energy conservation dictates that $\omega_A = \omega_3$. In other words, the energy of the incoming field $\phi$ should match the energy of the outgoing phonon mode. Similarly, for the second diagram, energy conservation requires that $\omega_1 + \omega_2 = \omega_3 + \omega_4$. This equation ensures that the total energy before and after the scattering process remains conserved.

In [13] the authors discussed the utilization of Kerr nonlinearity to mix waves, specifically external drives and phonon modes, for the implementation of quantum gates. In our approach we interpret these wave mixing processes as scattering processes represented by Feynman diagrams. The left vertex in FIG. 2 plays a crucial role in converting an incoming $\phi$ field into a phonon mode. Since $\phi$ is a scalar field, its propagator becomes a classical number, for $d = 1, 2, 3$. This gives rise to effective coupling with coupling constants $g_1$ and $g_2$.

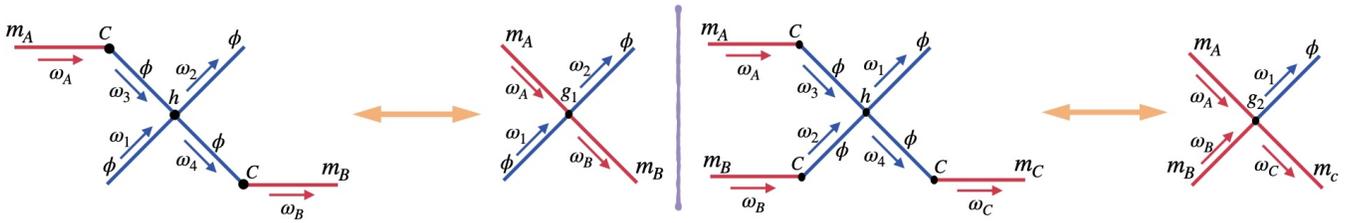

FIG. 3. The wave mixing process can be represented by tree-level Feynman diagrams in the Lagrangian density Eqn. (60). These diagrams involve the vertices specified in the Feynman rules of the Lagrangian density Eqn. (81). In these diagrams, energy conservation plays a crucial role and corresponds to the resonance approximation condition for implementing gate evolution operators with the Hamiltonian Eqn. (1), as discussed in [13]. The requirement of energy conservation ensures that the frequencies of the interacting modes satisfy specific conditions that enable efficient gate operations.

We emphasize that while the propagator of $\phi$ varies with different dimensions, the metric remains Lorentzian in all dimensions. As a result, the propagators can all be treated as classical numbers for $d = 1, 2, 3$. Consequently, the values of the coupling constants $g_1$ and $g_2$ may differ in different dimensions. Building on this observation, we construct an effective Lagrangian Eqn. (81), that incorporates the vertices representing the wave mixing processes and the corresponding coupling constants $g_1$ and $g_2$, as illustrated in FIG. 4. This is analogous to the construction of effective time evolution operators with effective coupling constants $g_v^{(1)}$ and $g_v^{(2)}$ in [13].

$$\mathcal{L} = \frac{\rho}{2}|\dot{\mathcal{U}}|^2 - \sum_{\alpha,\beta}^{d} \frac{\lambda^{(d)}}{2}\left(\frac{\partial \mathcal{U}_\alpha}{\partial x^\beta}\right)^2 + \frac{1}{2}\partial_\mu \phi \partial^\mu \phi - \frac{g_1}{2l_0^{2-d}}\sum_{\alpha \neq \beta} \phi^2 \mathcal{U}_\alpha \mathcal{U}_\beta - \frac{g_1}{4l_0^{2-d}}\sum_\alpha \phi^2 \mathcal{U}_\alpha^2$$
$$- \frac{g_2}{l_0^{2-d}}\sum_{\alpha \neq \beta \neq \gamma} \phi \mathcal{U}_\alpha \mathcal{U}_\beta \mathcal{U}_\gamma - \frac{g_2}{2l_0^{2-d}}\sum_{\alpha \neq \beta} \phi \mathcal{U}_\alpha^2 \mathcal{U}_\beta - \frac{g_2}{3!l_0^{2-d}}\sum_\alpha \phi \mathcal{U}_\alpha^3 \,. \tag{81}$$

Since we are looking at the same Lagrangian for different dimensions, to keep $g_1$ and $g_2$ of dimension $L^{-1}$, we add a unit length factor $l_0$ to the denominator. Conventionally, we use god-given units where Plank's constant $\hbar$ and the speed of light $c$ are set to be 1, such that the unit of time is the same as the unit for length. We note that in our analysis we are assuming that the couplings for different polarizations are the same, taking into account various combinatorical factors. This assumption is similar to the isotropic assumption while we constructed the system. This simplification allows us to treat the couplings uniformly and facilitates the analysis of the wave mixing processes and gate operations.

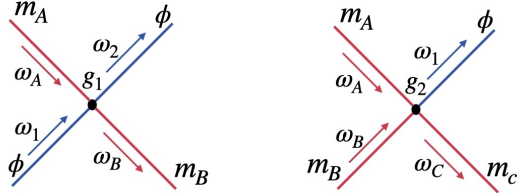

FIG. 4. Vertices in Feynman rules or basic couplings of Lagrangian density Eqn. (81).

Now, let us proceed with the calculation of clock cycle times in this quantum field theory (QFT) formulation. First, we note that the uncontracted phonon field operates on a phonon state in the following manner:

$$\langle 0|\mathcal{U}(x)|\omega_A\rangle = \int \frac{d^d\mathbf{k}}{(2\pi)^{d/2}\sqrt{2\rho\omega_\mathbf{k}}} \langle 0|\left(m_\mathbf{k} e^{-i\mathbf{k}\cdot\mathbf{x}-i\omega_\mathbf{k}t} + m_\mathbf{k}^\dagger e^{i\mathbf{k}\cdot\mathbf{x}+i\omega_\mathbf{k}t}\right)|\omega_A\rangle = e^{-i\mathbf{k}_A\cdot\mathbf{x}-i\omega_A t} \,. \tag{82}$$

The Feynman diagrams represent scattering matrix elements, such as for SWAP gates, where $\langle\omega_B|U(T)|\omega_A\rangle$ corresponds to the first diagram in FIG. 4. It describes a scattering process involving incoming states $|\omega_A\rangle$ and $|\omega_1\rangle$ and outgoing states $|\omega_B\rangle$ and $|\omega_2\rangle$. In the following analysis we will omit the $\phi$ states and operators since our focus is solely on the phonon modes involved in QRAM operations. Additionally, we will consider only the leading-order contributions, which in this case correspond to the two tree-level processes depicted in FIG. 4:

$$\langle\omega_B|U(T)|\omega_A\rangle = \langle\omega_B|\mathbf{T}\exp\left\{-i\int_{-t_{sw}/2}^{t_{sw}/2} dt \int d^d\mathbf{x}\, \frac{g_1}{4l_0^{2-d}}\phi^2\mathcal{U}^2\right\}|\omega_A\rangle$$
$$= \langle\omega_B|\exp\left\{ig_1 t_{sw}(m_A^\dagger m_B + m_B^\dagger m_A)\right\}|\omega_A\rangle \,. \tag{83}$$

In this particular Feynman diagram there are no internal lines, indicating the absence of contractions. Therefore, the time-ordered product is transformed into a normal-ordered product using Wick's theorem [5]. The combinatorial factors take into account the various possibilities of using different fields to annihilate or create particles. Consequently, the constant in front of the diagram is just $g_1$. Furthermore, note that we constantly ignore the overall delta functions which guarantees the energy-momentum conservation. This is the similar resonance dominant approximation that was used in [4].

The energy conservation condition for the SWAP gates, $\omega_A + \omega_1 = \omega_B + \omega_2$, ensures that all the time-dependent parts vanish. As a result, the integral over $t$ simplifies to a multiplication by $t_{sw}$ in the second line. It's important to note that we are asking a different question compared to traditional QFT. In QFT, calculating the scattering amplitude involves determining the possibility of a specific scattering process occurring. In this context, however, we are interested in determining how long the operators need to act on the initial states given a specific final state.

In the last line of Eqn. (83), we have obtained the unitary time evolution operator for the SWAP gates, which is exactly the same as the one obtained in [13]. The additional delta function accounts for momentum conservation. We denote this unitary operator as $S(t) = \exp ig_1 t(m_A^\dagger m_B + m_B^\dagger m_A)$.

$$
\begin{aligned}
&\langle\omega_B|\exp\left\{ig_1t_{sw}(m_A^\dagger m_B + m_B^\dagger m_A)\right\}|\omega_A\rangle \\
&= \langle\omega_B|\exp\left\{ig_1t_{sw}(m_A^\dagger m_B + m_B^\dagger m_A)\right\}m_A^\dagger|0\rangle \\
&= \langle\omega_B|\exp\left\{ig_1t_{sw}(m_A^\dagger m_B + m_B^\dagger m_A)\right\}m_A^\dagger \exp\left\{-ig_1t_{sw}(m_A^\dagger m_B + m_B^\dagger m_A)\right\}|0\rangle \\
&= \langle\omega_B|\left(m_A^\dagger \cos(g_1t_{sw}) + im_B^\dagger \sin(g_1t_{sw})\right)|0\rangle .
\end{aligned} \quad (84)
$$

The third line is obtained by assuming that we are dealing with a two-state system. We insert $1 = S(t)^\dagger S(t)$, where the operator $S(t)$ acts trivially on the vacuum state. By setting the SWAP time scale as $t_{sw} = \frac{\pi}{2g_1}$, we ensure that the final state is $|\omega_B\rangle$. The universal phase factor $i$ in front is not important for our purposes.

The beam-splitter gate is realized by using the same energy conservation condition as the SWAP gates. To obtain the time scale for the beam-splitter, we need to modify the final state in the calculation given in Eqn. (84):

$$
\begin{aligned}
\frac{1}{\sqrt{2}}(\langle\omega_A| + i\langle\omega_B|)\exp\{ig_1t_{bs}(m_A^\dagger m_B + m_B^\dagger m_A)\}|\omega_A\rangle &= 1 , \\
\cos(g_1t_{bs}) = \sin(g_1t_{bs}) &= \frac{\pi}{4} .
\end{aligned} \quad (85)
$$

Therefore, the time for a beam-splitter is given by $t_{bs} = \frac{\pi}{4g_1}$. To realize CZ gates, we need to consider the other type of wave mixing shown in FIG. 3. The energy conservation condition for CZ gates is $\omega_A + \omega_B = \omega_C + \omega_1$. The corresponding evolution operator for the CZ gate is

$$
\langle\omega_A,\omega_B|\exp\{ig_2t_{cz}(m_A^\dagger m_B^\dagger m_C + m_C^\dagger m_A m_B)\}|\omega_A,\omega_B\rangle = -1 . \quad (86)
$$

A calculation similar to that in Eqn. (84) will give the time for CZ gate: $t_{cz} = \pi/g_2$. Note that all three unitary operators and corresponding time scales are in the same form as calculated in [13] using Hamiltonian Eqn. (1). We present a full list of the correspondence of the Feynman diagrams of Eqn. (81) and quantum gates to be used for implementing QRAM in FIG. 5. Note that the beam-splitter has the same Feynman diagram with the SWAP gate. As shown above, the only difference between them is the time scale.

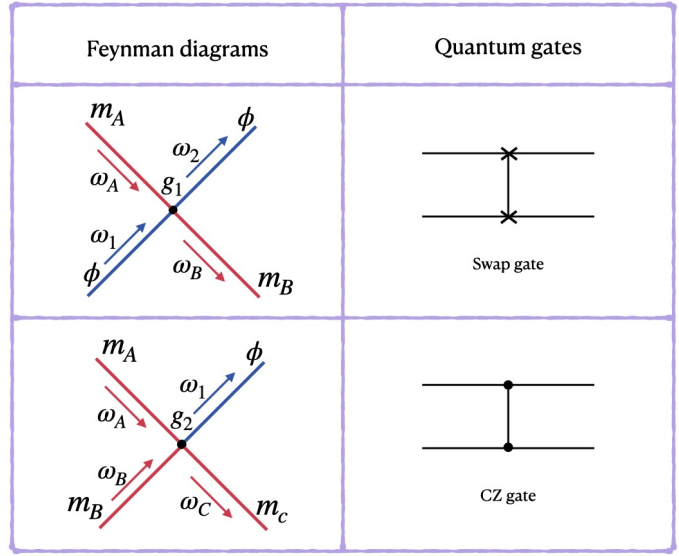

FIG. 5. Correspondence between the Feynman diagrams of Eqn.(81) and the quantum gates used for implementing QRAM.

Moreover, in the work by [13], corrections to the two effective coupling constants $g_1$ and $g_2$ were considered through perturbation theory. Similarly, in the QFT formulation, we can calculate more complex Feynman diagrams involving the same initial

and final states to obtain higher-order corrections to $g_1$ and $g_2$. This includes the possibility of loop diagrams, which can lead to the renormalization of the coupling constants. The renormalization theorem discussed in [5] is useful here for determining the renormalizability of a QFT.

**Theorem II.1.** *A Lagrangian involving all interaction of raw dimension less than or equal to 4 is strictly renormalizable.*

Both the interactions described by Eqn. (81) and Eqn. (60) have raw dimensions of exactly 4. This indicates that both theories are strictly renormalizable, meaning that the divergences encountered in quantum field theory can be systematically removed through renormalization procedures. Therefore, there is no concern about the theory going to infinity.

## III. BOUNDS OF QRAM BY LOCALITY

In the preceding sections, we derived the locality constraint on information propagation in a QRAM system using both discrete lattice and continuous field theory constructions. This constraint is imposed by the Hamiltonian described by Eqn. (19) and Eqn. (60) (under the assumption of isotropy). As a result of these constraints, there is a bound on the size of the QRAM system. The information within the system, governed by Eqn. (60), can propagate at most a distance of $N \times a$, where $N$ is the total number of qubits and $a$ is their separation, through a series of quantum gates generated by the same equation. The total time required for these operations represents the overall time scales. Therefore, we can interpret the speed at which information travels to the output as the distance $N \times a$ divided by the total time. However, it is crucial to ensure that this speed does not exceed the bound set by Eqn. (32) or Eqn. (80).

$$\frac{\text{Total distance information travel}}{\text{Total operation time}} \leq \text{Speed limit for information propagation} . \tag{87}$$

The total operation time in the QRAM system is indeed related to the depth of the QRAM, which can be thought of as layers of quantum routers. These quantum routers serve as the building blocks of the QRAM, responsible for routing phonon modes to the appropriate channels for information extraction or for routing information out of the system. The depth of the QRAM corresponds to the number of these layers, which is given by $\log N$, where $N$ represents the total number of qubits. Therefore, the clock cycle time required for the operations in the QRAM system scales with the logarithm of the number of qubits.

### A. Total clock cycle time for QRAM

In this subsection we will determine the total operation time required for extracting both classical and quantum information from a QRAM system of size $N$, using the QRAM construction outlined in [13]. The system consists of two fundamental steps: initialization and routing. The initialization step is performed by using a SWAP gate, while the routing step involves a combination of a controlled-SWAP gate and a SWAP gate, as depicted in FIG. 6.

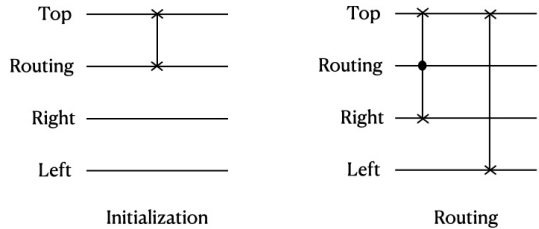

FIG. 6. Initialization refers to the process of exchanging the address qubit, located at the "top" channel, with the routing qubit. The routing step consists of two operations: a controlled-SWAP gate with the right channel and a swap operation with the left channel. The behavior of the routing step depends on the state of the routing qubit. If the routing qubit is in the state $|1\rangle$, the "top" channel is swapped with the right channel, while the swap operation exchanges two $|0\rangle$ states. However, if the routing qubit is in the state $|0\rangle$, the controlled-SWAP gate has no effect, and the swap operation swaps the "top" channel with the left channel.

The controlled-SWAP gate can be constructed by using two beam-splitter gates and a CZ gate. The gate operation $U^{(123)}csw$ is composed of the following sequence: first, we apply the beam-splitter gate $Ubs^{(23)}$ to channels 2 and 3, then the CZ gate $U_{cz}^{(12)}$ is applied to channels 1 and 2, and finally, we apply the beam-splitter gate $U_{bs}^{\dagger(23)}$ to channels 2 and 3. This sequence of operations implements the controlled-SWAP gate, with channel 1 (the routing qubit) remaining unchanged during the process.

$$\begin{aligned}
\mathcal{U}_{csw}^{(123)}\left|1,1,0\right\rangle &= \mathcal{U}_{bs}^{\dagger(23)}\mathcal{U}_{cz}^{(12)}\mathcal{U}_{bs}^{(23)}\left|1,1,0\right\rangle \\
&= \mathcal{U}_{bs}^{\dagger(23)}\mathcal{U}_{cz}^{(12)}\frac{1}{\sqrt{2}}(\left|1,1,0\right\rangle + i\left|1,0,1\right\rangle) \\
&= \mathcal{U}_{bs}^{\dagger(23)}\frac{1}{\sqrt{2}}(-\left|1,1,0\right\rangle + i\left|1,0,1\right\rangle) \\
&= -\frac{1}{2}(\left|1,1,0\right\rangle - i\left|1,0,1\right\rangle) + \frac{1}{2}(i\left|1,1,0\right\rangle + \left|1,0,1\right\rangle) \\
&= i\left|1,0,1\right\rangle \ .
\end{aligned} \tag{88}$$

The combination of gates to form the controlled-SWAP gate, as described above, is presented in FIG. 7. Please note that the overall phase factor in the gate representation can be ignored for our purposes.

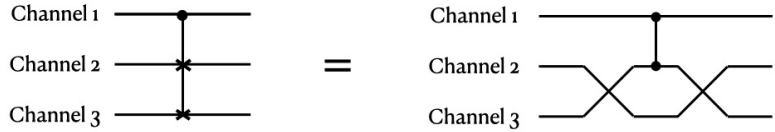

FIG. 7. A controlled-SWAP gate can be constructed using a beam splitter on channels 2 and 3 (the first cross), a CZ gate on channels 1 and 2, and another beam-splitter on channels 2 and 3 (the second cross). Channel 1 serves as the control qubit and remains unchanged throughout the gate operation. This construction is illustrated in the diagram provided in FIG. 7. The gate operation can be represented by the equation given in Eqn. (88).

It's worth noting that the decomposition of controlled-SWAP gates shown in FIG. 7 can work for more general cases. While we assumed all three channels are qubits, it is only necessary for channel 1 to be qubits, and channels 2 and 3 can be bosonic modes to make this decomposition work. Moreover, under the assumption of three qubits, there might be even simpler decompositions of controlled-SWAP gates, such as using three Toffoli gates.

For experimental demonstrations of this composite design of controlled-SWAP gates, reference [14] discusses controlled-SWAP gates using the same circuit in FIG. 7, but they did not demonstrate the coherence of controlled-SWAP due to significant dephasing in channel 1. However, reference [15] uses a slightly different method and improved device coherence properties to demonstrate deterministic controlled-SWAP operation.

In the Bucket-brigade QRAM architecture, the initialization of the address qubits is performed sequentially, one by one. This means that each address qubit is routed into the system for initialization individually. In an $N$-qubit QRAM, it would take $\log N$ steps to fully initialize all the qubits. This sequential routing and initialization process can be visualized as having $\log N$ layers of quantum routers, hence the term "QRAM depth."

In each step $k$, as depicted in FIG. 8, the $k$th address qubit is first routed (or swapped) $k-1$ times, passing through the previous layers of quantum routers. After being routed, the qubit undergoes a swap operation to perform the initialization. This process is repeated for each address qubit, following the sequential order.

This sequential routing and initialization scheme allows for the efficient and controlled initialization of the qubits in the QRAM system, ensuring that each qubit is properly addressed and prepared for subsequent operations.

Hence, to fully initialize the QRAM system, we need to apply SWAP gates (for initialization) $\log N$ times and a combination of controlled-SWAP and SWAP gates (for downstream routing) $\log N - 1$ times.

The process of extracting classical information from the QRAM system depends on the specific form of the information that needs to be extracted. After routing the bus qubit $\log N$ times, we can copy the information from the database by applying a CZ gate. Subsequently, we can route out the bus qubit and all the address qubits from the system to disentangle them. These extraction steps take the same amount of time as the full initialization of the system. We denote the total operation time for extracting classical information as $T_c$.

$$\begin{aligned}
T_c =& 2 \times \frac{(1+\log N)\log N}{2} \times t_{sw} + 2 \times \frac{(\log N - 1)\log N}{2} \times (2t_{bs} + t_{cz}) \\
& + 2\log N \times (2t_{bs} + t_{cz} + t_{sw}) + t_{cz} \\
=& \left(\log^2 N + 2\log N\right)\frac{\pi}{g_1} + \left(\log^2 N + \log N + 1\right)\frac{\pi}{g_2} \ .
\end{aligned} \tag{89}$$

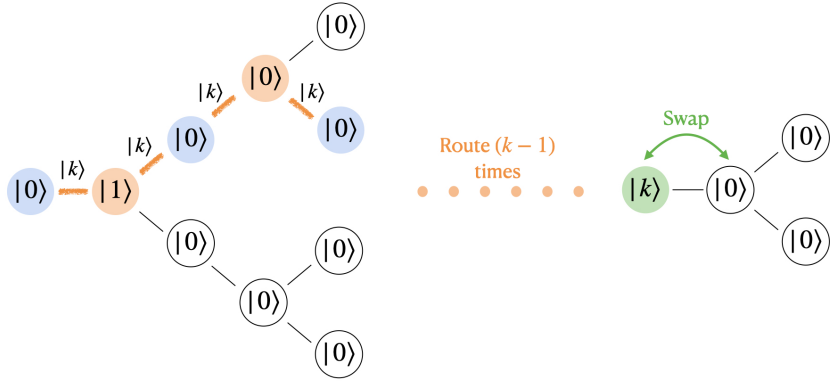

FIG. 8. During the $k$th step of the initialization process, the $k$th address qubit $|k\rangle$ is routed through the quantum routers. In the figure, the routing path for $|k\rangle$ is represented by the yellow branches, connecting the blue sites from left to right for each quantum router. In this particular system, the $|k\rangle$ qubit is first routed from the first blue site to the second blue site, bypassing the white qubit below, because the first routing qubit (yellow) is in state $|1\rangle$. Then, it is further routed to the third blue site because the second routing qubit is in state $|0\rangle$. Finally, the $|k\rangle$ qubit reaches the green site, which corresponds to the "top" channel of the quantum router at depth $k$. At this point, a swap operation is performed between $|k\rangle$ and the routing qubit to complete the initialization process.

The first term corresponds to the SWAP gates used for initialization, while the second term represents the routing steps involved in the initialization process. The third term accounts for the routing of the bus qubit, and the last term refers to the information extraction process. The first three terms are all applied 2 times for upstream routing (bus qubit and address qubits) and disentanglement, which is stage 4 and 5 in [1].

In the case of quantum information, we cannot directly copy the information due to the no-cloning theorem. Instead, we route a pointer qubit $\log N$ times to the appropriate location and utilize a controlled-SWAP gate to extract the information [13]. Subsequently, we need to route out both the information state, the pointer qubit, and all the address qubits from the system. The total operation time for quantum information, denoted as $T_q$, is the sum of all the preceding steps (full initialization and routing of the pointer qubit) plus an additional $\log N$ times for routing the information qubit:

$$\begin{aligned} T_q =& 2 \times \frac{(1+\log N)\log N}{2} \times t_{sw} + 2 \times \frac{(\log N - 1)\log N}{2} \times (2t_{bs} + t_{cz}) \\ &+ 2\log N \times (2t_{bs} + t_{cz} + t_{sw}) + 2t_{bs} + t_{cz} + (\log N - 1)(2t_{bs} + t_{cz}) \\ =& \left(2\log^2 N + 5\log N\right)\frac{\pi}{2g_1} + \left(\log^2 N + 2\log N\right)\frac{\pi}{g_2} \, . \end{aligned} \qquad (90)$$

Therefore, in the case where $N$ is significantly larger than 1, the leading-order approximation of the total operating time for transferring classical and quantum data is proportional to the square of the depth. Specifically, the total operation time $T_{c\text{ or }q}$ can be expressed as

$$T_{c\text{ or }q} \sim \log^2 N \left(\frac{\pi}{g_1} + \frac{\pi}{g_2}\right). \qquad (91)$$

This implies that the total operation time for extracting both classical and quantum information scales with the square of the logarithm of $N$. We note that this specific result may not hold in all cases, as alternative approaches such as fan-out QRAM can be used [1]. However, regardless of the specific QRAM implementation, the operation time is generally expected to be proportional to $\log N$ since there are typically $\log N$ layers involved.

### B. Limit on the size of QRAM

With the total extraction time $T_{c\text{ or }q}$ given by Eqn. (91), we can now provide a more quantitative understanding of Eqn. (87). The basic idea is that the time required to extract information from a QRAM system scales with the logarithm of the system size

$N$. This scaling behavior arises from the inherent depth of the QRAM, which consists of multiple layers of quantum routers.

In particular, Eqn. (87) states that the time complexity of information extraction from the QRAM system is proportional to $\log^2 N$. This indicates that as the size of the system increases, the total extraction time grows quadratically with the logarithm of $N$. The coefficients in front of $\log^2 N$, namely, $\frac{\pi}{g_1} + \frac{\pi}{g_2}$, represent the contribution of the specific QRAM implementation and the associated coupling constants:

$$\frac{N}{\log^2 N} \leq \frac{4\sqrt{d}\left(\frac{\pi}{g_1} + \frac{\pi}{g_2}\right)}{a} \left(\sum_{j=1}^{\nu} \frac{\lambda_j}{m}\right)^{1/2} \quad \text{or} \quad \left(\frac{\lambda^{(d)}}{\rho}\right)^{1/2}. \tag{92}$$

In FIG. 9 we illustrate the relationship between the material constants used in the construction of the QRAM and the corresponding size limit imposed by the locality constraint.

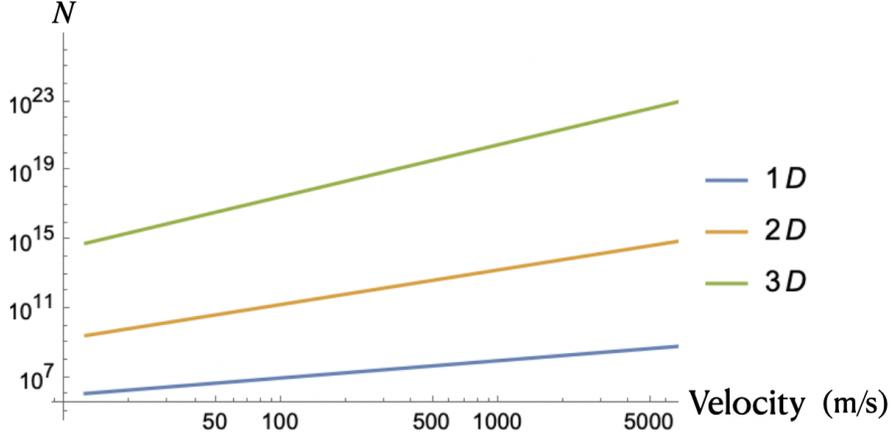

FIG. 9. Bounds of QRAM size $N$ for dimensions 1, 2, and 3. Bounds of QRAM size $N$ for dimensions 1, 2, and 3. Here we assume the lattice spacing of $10^{-6}$ m and the clock cycle time of $10^{-3}$ s. The horizontal axis is equal to the velocity limit determined by $\sum_{j=1}^{\nu} \sqrt{d}\left(\frac{\lambda_j}{m}\right)^{1/2}$ or $\sqrt{\frac{\lambda^{(d)}}{\rho}}$. This velocity is taken to be at most on the order of typical sound speed in solids: about 6000 m/s.

Please note that in our analysis, we have assumed a clock cycle time of $10^{-3}$ seconds, which is not necessarily the typical time used in the model presented in [13]. The choice of clock cycle time might be influenced by the use of perturbation theory in the construction, which requires the coupling constants $g_1$ and $g_2$ to be relatively small (less than 1).

Additionally, we have considered the speed limit to be the typical sound speed in solid materials. However, one can design alternative QRAM systems that can exceed this speed limit. In such cases, the ultimate speed limit would be determined by the speed of light, which is approximately $3 \times 10^8$ m/s (as indicated in Eqn. (65)).

An example of an alternative architecture is designed in [11], which utilizes quantum teleportation for qubit routing in QRAM instead of swap actions. In this system, actions based on Bell state measurement or teleportation are constrained by the speed of light, while other actions such as information extraction and controlled-SWAP actions are confined by the speed of sound.

As the number of qubits, $N$, increases, more actions are confined by the speed of light. For instance, in the case of $N = 4$, the entire system is confined by the speed of sound. For $N = 16$, however, QRAM operations involve longer distances to be routed through quantum teleportation. Consequently, if $N$ becomes very large, a significant portion of the routing will be dedicated to teleportation, while only a small part will be confined by the speed of sound. The actions confined by the speed of sound will occur mainly within the $N = 4$ subpart, involving controlled-SWAP operations and information extraction.

Roughly speaking, we estimate that around $2/N^{1/2}$ of the total actions should be constrained by the speed of sound, while the remaining portion, approximately $1 - 2/N^{1/2}$, will be constrained by the speed of light. We note that since this system is two-dimensional, the length scale of the system is $N^{1/2} \times a$, where $a$ is the qubit separation. We assume a time scale of approximately $10^{-3}$ and a qubit separation of roughly 1 micrometer.

$$\frac{N^{1/2} \times a}{T \log N^{1/2}} \leq \left(\frac{2}{N^{1/2}}\right) \times 6 \times 10^3 + \left(1 - 2/N^{1/2}\right) \times 3 \times 10^8. \tag{93}$$

Hence, the QRAM size $N$ of this 2D design ranges from approximately $\mathcal{O}(10^{20}) \sim \mathcal{O}(10^{22})$ qubits. This range lies above the yellow line (representing 2D systems) in FIG. 9 but below the green line (representing 3D systems) at $\sum_{j=1}^{\nu} d(\lambda_j/m) = 3.6 \times 10^7$ m/s.

Nevertheless, we note that even when considering the bound to be the sound speed, our results demonstrate that the maximum bounds on the size of QRAM systems might lie in the large scale. For example, in the case of a 1D system, our findings indicate that the bounds on $N$ reaches nearly $10^9$ qubits. Moreover, if we consider systems in two or three dimensions, the bounds on $N$ are even higher, which ensures that these bounds might be weak enough for some quantum algorithms.

On the other hand, by considering both $g_1$ and $g_2$ to be approximately equal, we have created another plot in FIG. 10 that depicts this limit. The plot is presented as a heat plot, with the coupling constants $g_1$ and $g_2$ represented on the x-axis and the square of the speed limit on the y-axis.

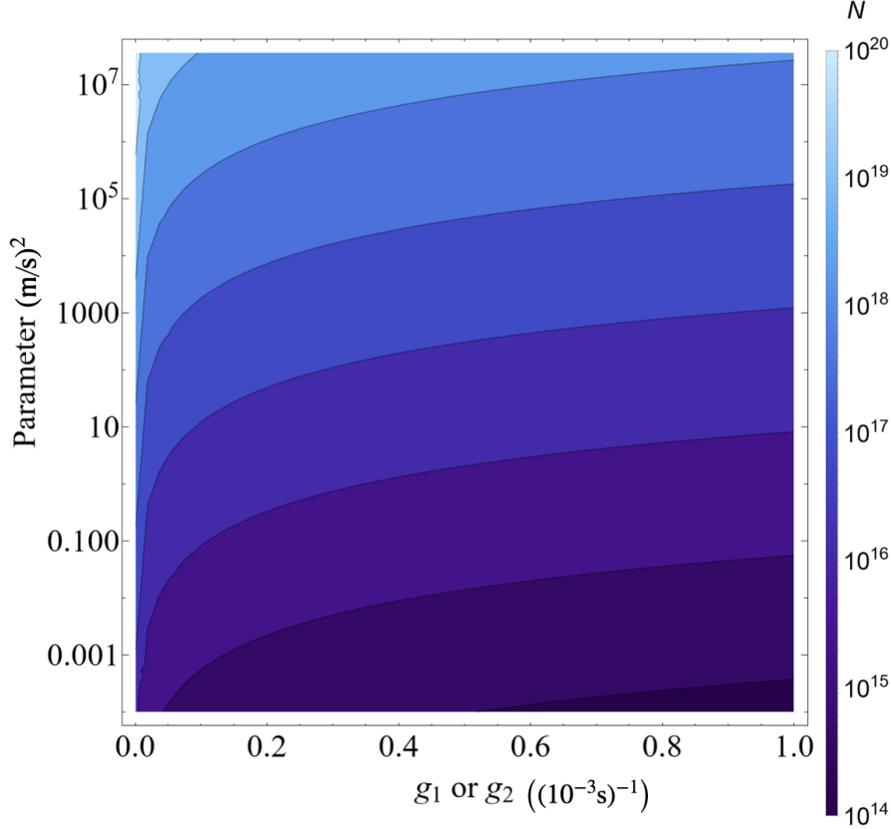

FIG. 10. We have chosen $g_1$ and $g_2$ to be less than 1 due to our utilization of perturbation theory throughout the construction process. If either of the coupling constants is not sufficiently small or weakly coupled, the fundamental assumptions of perturbative quantum field theory no longer hold. In such cases, the leading order is determined by the Feynman diagram with the greatest number of vertices. The range of $\sum_{j=1}^{\nu} \lambda_j$ in the plot is consistent with that in FIG. 9. For parameters in the vertical axis we indicate $\sum_{j=1}^{\nu} \left(\frac{\lambda_j}{m}\right)$ or $\frac{\lambda^{(1d)}}{\rho}$.

This plot demonstrates that for the 1D case, when considering a more realistic time scale determined by the coupling constants $g_1$ and $g_2$, the limit on the size of the QRAM ($N$) remains quite large (reaching $10^{14}$ qubits for 1 dimension). However, it is important not to go too far to the left in FIG. 10, as that would result in excessively long waiting times for QRAM operations.

Thus, the larger the quantity $\left(\sum_{j=1}^{\nu} \lambda_j\right)$ is, the larger the system can be. On the other hand, reducing the coupling constants $g_1$ and $g_2$ allows for larger systems, but it also leads to longer clock cycle times. To increase the upper limit, one might need to explore materials where atoms are strongly coupled or reduce the coupling constants between the transmon qubit and the resonator to increase the clock cycle time. However, we note that the upper bounds presented in FIG. 9 and FIG. 10 are already sufficient for the requirements of most quantum algorithms that involve a large number of qubits.

## C. Hybrid designs of QRAM

The time scale $\Delta T$ used for the clock cycle time is closely related and bounded by the gate time and the decoherence time of qubits. Reducing $\Delta T$ will lead to more stringent causality bounds for QRAM. However, this does not mean that we should make $\Delta T$ as large as we can in order to snake around the bound. In fact, smaller $\Delta T$ leads to faster gates and more efficient operations

of QRAM circuits. Moreover, according to [16], smaller $\Delta T$ will lead to smaller infidelities and better error resilience of QRAM architectures.

Here is an example showing that our QRAM causality bound is compatible and friendly with hybrid designs of QRAM architecture. According to the inequality Equ.(1) in the main text in our one-dimensional example,

$$\frac{L}{T} = \frac{Na}{\Delta T \log N} \le c \,. \tag{94}$$

The value of $N$ saturating the bound, $N_*$, could be estimated as,

$$N_* \approx \frac{c\Delta T}{a} \,, \tag{95}$$

since $\log N$ could be ignored compared to $N$ at large $N$. On the other hand, one can estimate the infidelity $1 - F \propto \varepsilon \Delta T$, where $\varepsilon$ is the error rate per unit time. Thus, one could choose an optimal $\Delta T$ by both considering infidelity and the causality bound, by minimizing a loss function,

$$L = a_1 \frac{1}{\varepsilon \Delta T} + a_2 \frac{c\Delta T}{a} \,, \tag{96}$$

for some positive dimensionless constants $a_{1,2}$. Thus, one could choose an optimal $\Delta T$,

$$\Delta T_* \sim \sqrt{\frac{aa_1}{c\varepsilon a_2}} \,. \tag{97}$$

This analysis demonstrates the possibility of hybrid designs considering causality bounds and other factors.

---


}

\end{document}